\documentclass[twocolumn,prd,floatfix,preprintnumbers,a4paper,nofootinbib,superscriptaddress,groupedaddress]{revtex4-1}
\usepackage[utf8]{inputenc}
\usepackage{amsmath,amssymb}
\usepackage{amsfonts}
\usepackage{bm}
\usepackage{courier}
\usepackage{graphics}
\usepackage{float}
\usepackage{bm} 
\usepackage{dcolumn}
\usepackage{color}
\usepackage{mathrsfs}
\usepackage{textcomp}
\usepackage{amsmath}
\usepackage{amssymb}
\usepackage[normalem]{ulem} 
\usepackage[mathscr]{euscript}
\usepackage{acronym}
\usepackage{float}
\usepackage{varioref}
\usepackage[caption=false]{subfig}
\usepackage{lipsum}
\usepackage{mathtools}
\usepackage{multirow}
\usepackage{graphicx}
\usepackage{mathtools}
\usepackage{multirow}
\usepackage{graphicx}
\usepackage[usenames,dvipsnames,table,xcdraw]{xcolor}
\usepackage[colorlinks, pdfborder={0 0 0}]{hyperref} 
\usepackage[nameinlink,capitalise]{cleveref}
\usepackage{tensor}
\usepackage{verbatim}
\usepackage{wrapfig}
\usepackage{soul} 

\begin{document}

\title{Implications of the quantum nature of the black hole horizon\\
on the gravitational-wave ringdown}

\author{
		Sumanta Chakraborty$^{1}$,
		Elisa Maggio$^{2,3}$,
		Anupam Mazumdar$^{4}$,
		Paolo Pani$^{2}$.
	}
	\affiliation{$^{1}$School of Physical Sciences, Indian Association for the Cultivation of Science, Kolkata-700032, India}
	\affiliation{$^{2}$Dipartimento di Fisica, ``Sapienza'' Universit\`a di Roma \& Sezione INFN Roma1, P.A. Moro 5, 00185, Roma, Italy}
	\affiliation{$^{3}$Max Planck Institute for Gravitational Physics (Albert Einstein Institute)\\
Am Mühlenberg 1, 14476 Potsdam, Germany}
	\affiliation{$^{4}$Van Swinderen Institute, University of Groningen, 9747 AG, Groningen, The Netherlands}

\begin{abstract}

Motivated by capturing putative quantum effects at the horizon scale, we model the black hole horizon as a membrane with fluctuations following a Gaussian profile. By extending the membrane paradigm at the semiclassical level, we show that the quantum nature of the black hole horizon implies partially reflective boundary conditions and a frequency-dependent reflectivity.
This generically results into a modified quasi-normal mode spectrum and the existence of echoes in the postmerger signal. 
On a similar note, we derive the horizon boundary condition for a braneworld black hole that could originate from quantum corrections on the brane. This scenario also leads to a modified gravitational-wave ringdown. We discuss general implications of these findings for scenarios predicting quantum corrections at the horizon scale.
\end{abstract}
\maketitle
\section{Introduction}

According to the standard ``no drama" scenario, nothing unusual should happen when an object falls through the event horizon of a classical black hole~(BH), since locally the horizon does not represent any special region of spacetime. However, this picture might lead to contradictions when accounting for quantum effects, as dramatically put forward by the information loss paradox~\cite{Hawking:1975vcx,Mathur:2009hf,Polchinski:2016hrw,Chakraborty:2017pmn,Raju:2020smc}. 
In the attempts to solve these problems, the BH horizon acquires a special role as it sets the scale in which quantum effects might become important regardless of the curvature scale of the object~\cite{Mathur:2005zp,Mathur:2008nj,Mathur:2009hf,Visser:2009pw,Barcelo:2010xk,Giddings:2017jts,Carballo-Rubio:2018jzw,Koshelev:2017bxd,Buoninfante:2018rlq,Buoninfante:2018xiw,Buoninfante:2019teo}.
If this is the case, astrophysical BHs might provide a unique portal to quantum gravity phenomenology~\cite{Cardoso:2016oxy,Cardoso:2016rao,Abedi:2016hgu,Holdom:2016nek,Abedi:2020ujo,Maggio:2020jml,Maggio:2021ans,Addazi:2021xuf}.

Providing a concrete model for the quantum structure of the horizon is challenging and often based on vague proposals (see~\cite{Burgess:2018pmm,Oshita:2019sat} for some proposals to include quantum effects at the horizon).
In this work, in an attempt to quantify the effect of a quantum structures near the horizon, we model the quantum properties of BHs in two complementary ways.

First, we make use of the BH membrane paradigm~\cite{Damour:1982,Thorne:1986iy,Price:1986yy} and recent extensions thereof~\cite{Oshita:2019sat,Maggio:2020jml} to model the quantum properties of the horizon with a semiclassical membrane, such that the (classical) background spacetime is perturbed by the quantum stress-energy tensor of the membrane.
A second complementary possibility is to consider a modified BH background (possibly due to quantum effects) and apply the classical membrane paradigm by replacing the modified horizon with a classical fictitious membrane.
In the first scenario, the background is classical and the perturbation is of quantum origin and thus the coupling between geometry and matter is treated semiclassically; whereas, in the second scenario, the background has a quantum origin (at least semiclassically) and the perturbation is classical.
The latter possibility is realized, for instance, in the braneworld scenario~\cite{Maartens:2003tw,Csaki:2004ay,Perez-Lorenzana:2005fzz,Randall:1999ee,Arkani-Hamed:1998jmv,Harko:2004ui,Dadhich:2000am,Chamblin:1999by}, where the BH localized on the brane necessarily harbours quantum corrections from the bulk spacetime.
In this context, the vacuum solution on the brane does not correspond to a Schwarzschild BH, rather it resembles a tidally charged BH. The value of the tidal charge is related to the size of the extra dimensions by extending the brane geometry to the bulk spacetime~\cite{Chamblin:2000ra,Emparan:1999wa}.

The first and second scenarios are discussed in Sec.~\ref{sec:quantummembrane} and Sec.~\ref{sec:braneworld}, respectively. In each section, we derive the boundary conditions on the modified horizon due to quantum effects. We compute the quasi-normal mode (QNM) spectrum and the gravitational-wave (GW) echoes in the postmerger phase~\cite{Cardoso:2016oxy,Cardoso:2016rao,Abedi:2016hgu,Conklin:2017lwb,Cardoso:2017cqb} (see~\cite{Cardoso:2019rvt,Abedi:2020ujo,Maggio:2021ans} for recent reviews).
A discussion and concluding remarks are given in Sec.~\ref{sec:conclusions}.
Hereafter we set the fundamental constants $G$ and $c$ to unity. Greek indices denote the spacetime coordinates and Latin indices denote the coordinates on the membrane hypersurface.

\section{Quantum membrane paradigm}\label{sec:quantummembrane}

According to the original membrane paradigm~\cite{Damour:1982,Thorne:1986iy,Price:1986yy}, a static observer can replace the interior of a BH by a fictitious membrane located at the horizon. The fictitious membrane is described by a viscous fluid whose physical properties are such that the membrane has the same phenomenology of the BH.
Recently, the membrane paradigm has been extended 
to the case of horizonless compact objects~\cite{Oshita:2019sat,Maggio:2020jml,Chen:2020htz,Xin:2021zir} to derive their GW signatures.

In this section, we shall extend the calculations presented in Ref.~\cite{Maggio:2020jml} by replacing the classical BH horizon with a fictitious quantum membrane. The fluid living on the membrane is made up of several quantum degrees of freedom. For the purpose of this discussion, we assume the quantum degrees of freedom to be in their ground state. All the physical quantities related to the fluid can be interpreted in terms of operators and thus enter in the classical discussion through their expectation values. In what follows, we first set up the basic features of the model and then discuss the background geometry and the gravitational perturbations thereof. Finally, we analyse the frequency-dependent reflectivity and show the QNM spectrum and the GW echoes for the quantum membrane. 

\subsection{The basic picture}\label{basic}

Quantum fluctuations at the horizon scale result in the presence of a fictitious membrane located outside the horizon. 
The fluid living on the membrane has a proper distance from the horizon that is related to the ground state of the quantum membrane. In the limit of a classical fluid, the proper distance of the fictitious membrane from the horizon is negligible and we retrieve the classical BH picture. Let us notice that, within our framework, it is also possible to consider the quantum fluid to be fluctuating around a classical surface close to, but not coincident with, the horizon.
In this case, in the limit of vanishing quantum corrections, we recover the result presented in Ref.~\cite{Maggio:2020jml} for horizonless compact objects (further details can be found in Appendix \ref{AppA}).

For simplicity, we shall focus on a static and spherically symmetric background geometry whose line element is given by
\begin{align}\label{ext_metric}
ds^{2}=-f(r)dt^{2}+\frac{1}{f(r)} dr^2+r^{2} \left(d\theta^2 + \sin^2 \theta d\phi^2\right) \,,
\end{align}
where $(t,r,\theta,\phi)$ are the usual Schwarzschild-like coordinates, and $f(r)$ is a  function of the radial coordinate such that one of its zeros is located at $r=r_{+}$, denoting the location of the horizon. 
We stress that the backreaction due to quantum effects on the background solution is assumed to be negligible, so effectively the background metric can be taken to be the Schwarzschild solution, i.e. $f(r)=1-2M/r$ with $M$ being the object's mass.

As described above, the quantum membrane is built on an ensemble of microscopic degrees of freedom in the ground state and is subjected to a harmonic oscillator potential. The position of the quantum degrees of freedom is depicted collectively by the operator $\hat{\epsilon}$, that plays the role of the position operator in quantum mechanics. At variance with the standard quantum mechanics case where the position vector can have continuous eigenvalues $\in(-\infty,+\infty)$, in the present context the eigenvalues of $\hat{\epsilon}$ $\in(0,\infty)$. This condition guarantees that the quantum membrane resides on (or outside of) the horizon and the probability of the quantum fluid to be inside the horizon is exponentially suppressed (effectively, zero).

The quantum state of the microscopic degrees of freedom, $|\Psi\rangle$, in the representation of $\hat{\epsilon}$ is $\Psi(\epsilon)=\langle \epsilon |\Psi\rangle$, where $|\epsilon \rangle$ are the eigenvectors of $\hat{\epsilon}$. Since the quantum degrees of freedom are in the ground state of a harmonic oscillator, the overall state of the quantum fluid is given by a Gaussian wave function,
\begin{align}\label{wave_function}
\Psi(\epsilon)=A\exp\left(-\frac{\epsilon^{2}}{2\sigma^{2}}\right) \,,
\end{align}
where $\epsilon$ is the eigenvalue of the operator $\hat{\epsilon}$ denoting the departure of the quantum membrane from the classical location of the horizon, and $A$ and $\sigma^2$ are the amplitude and the variance of the quantum state, respectively\footnote{The variance of the wave function captures the quantum nature of the system since $\sigma=\sqrt{\hbar/({\cal M}w)}$, where $\cal{M}$ and $w$ are the characteristic mass and frequency scales of the quantum degrees of freedom associated with the membrane. Thus, the classical limit, $\hbar \rightarrow 0$, is equivalent to $\sigma\rightarrow 0$.
}.
The normalization of the wave function requires
\begin{align}
|A|^{2}=\frac{2}{\sigma\sqrt{\pi}} \,.
\end{align}
%
The classical location of the membrane is at
\begin{align} \label{R}
R = r_{+}+\langle \hat{\epsilon}\rangle \,,
\end{align}
where the departure of the membrane from the location of the horizon is given by the expectation value of the operator $\hat{\epsilon}$ in the ground state $\Psi(\epsilon)$, i.e.,
\begin{align}\label{exp_epsilon}
\langle \hat{\epsilon}\rangle&=\int_{0}^{\infty} d\epsilon~ \epsilon |\Psi(\epsilon)|^{2}=\frac{\sigma}{\sqrt{\pi}}~.
\end{align}
Note that, in the limit of vanishing quantum corrections, the quantum nature of the membrane is absent and the classical membrane is located on the horizon. 
Finally, it is useful to determine the quantity $\langle \hat{\epsilon}^{2}\rangle-\langle \hat{\epsilon}\rangle^{2}$ that captures the quantum nature of the membrane in the most straightforward manner. From the wave function in Eq. (\ref{wave_function}), it follows that
\begin{align}\label{qdiff_caseII}
\langle \hat{\epsilon}^{2}\rangle-\langle \hat{\epsilon}\rangle^{2}&=\sigma^{2}\left(\frac{1}{2}-\frac{1}{\pi}\right)\,.
\end{align}
As we shall see, the presence of a nonzero $\sigma$ modifies significantly the BH boundary condition satisfied by the gravitational perturbations on the membrane. 

\subsection{The background geometry}

Having discussed the basic properties of the quantum membrane, let us focus on its effects on the background spacetime. The exterior geometry is described by the metric in Eq. (\ref{ext_metric}). The membrane satisfies the Israel-Darmois junction conditions~\cite{Darmois1927,Israel:1966rt,VisserBook} 
\begin{equation} \label{semi_class_junc}
[[K_{ab} - K h_{ab}]]=-8 \pi \langle \hat{T}_{ab} \rangle~, \qquad [[h_{ab}]]=0~,
\end{equation}
where $h_{ab}$ is the induced metric on the membrane, $K_{ab}$ is the extrinsic curvature, $K=K_{ab}h^{ab}$, $\langle \hat{T}_{ab}\rangle$ 
is the expectation value of the stress-energy tensor operator of the quantum matter distribution on the membrane, and $[[...]]$ is the jump of a quantity across the membrane.

According to the original membrane paradigm, the fictitious membrane is such that the extrinsic curvature of the internal spacetime vanishes, i.e., $K_{ab}^-=0$~\cite{Thorne:1986iy}, which we assume to be the case in the present context as well.
The junction conditions in Eq.~(\ref{semi_class_junc}) connect the classical geometry on the three-dimensional hypersurface located at $r=R$, to the quantum stress-energy tensor on the membrane, which we choose to be described by the following operator, 
\begin{align}\label{stress_energy}
\hat{T}_{ab}=\rho\hat{u}_{a}\hat{u}_{b}+\left(p-\zeta \hat{\Theta}\right)\hat{\gamma}_{ab}-2\eta \hat{\sigma}_{ab}~.
\end{align}
Here, $\hat u_{a}$ is the three-velocity of the fluid constituting the membrane, $\hat \sigma_{ab}$ is the associated shear tensor, $\hat{\Theta}$ is the expansion, and $\hat{\gamma}_{ab}$ is the metric induced on the two dimensional surface to which $\hat{u}_{a}$ is orthogonal. The definitions of the above quantities can be found in \cite{Maggio:2020jml}. 
It is to be noted that here these quantities are operators owing to their dependence on $\hat{\epsilon}$ through the location of the membrane.
To the first order in the gravitational perturbation, the energy density and the pressure of the fluid can be expanded as $\rho=\rho_{0}+\delta \rho$ and $p=p_{0}+\delta p$, where 
$\delta \rho$ and $\delta p$ are the first-order corrections (their expressions are in Appendix \ref{AppB} while discussing polar perturbations).

The parameters $\zeta$ and $\eta$ are the bulk and the shear viscosities governing how the fluid responds to the external perturbations. Exact computations of these parameters may arise from a quantum gravity model, which would also motivate the picture of the quantum membrane proposed here. In the absence of a consistent quantum gravity model, for simplicity, we  consider $\eta$ and $\zeta$ to be real constants; in particular, $\eta=1/(16\pi) \equiv \eta_{\rm BH}$ and $\zeta=-1/(16\pi) \equiv \zeta_{\rm BH}$ describe the classical BH case. Note that the viscosity parameters do not play any role in the unperturbed background geometry,  and hence the stress-energy tensor of the quantum membrane in the background is identical to that of a perfect fluid.

The background quantities
in the stress-energy tensor presented above can be computed as follows. From Eq.~(\ref{stress_energy}), the non-vanishing components of the stress-energy tensor are:
\begin{eqnarray}
\hat{T}_{tt} &=& \rho_{0}f(r_{+}+\hat{\epsilon}) \,, \\
\hat{T}_{\theta \theta} &=& p_{0}(r_{+}+\hat{\epsilon})^{2} = \frac{\hat{T}_{\phi \phi}}{\sin^{2}\theta} \,,
\end{eqnarray}
and the expectation values of the components of the stress-energy tensor are derived as\footnote{Note that these results can also be derived assuming a classical membrane fluid fluctuating following the distribution $|\Psi(\epsilon)|^{2}$.}
\begin{align}
\langle \hat{T}_{tt}\rangle&=\rho_{0}\langle f(r_{+}+\hat{\epsilon})\rangle
\nonumber
\\
&=\rho_{0} \left[f(R)+\frac{1}{2}f''(r_{+})\left(\langle \hat{\epsilon}^{2}\rangle-\langle \hat{\epsilon}\rangle^{2}\right)+\mathcal{O}\left(\tilde{\sigma}^{3}\right) \right]~,
\\
\langle \hat{T}_{\theta \theta}\rangle&=p_{0}\langle \left(r_{+}+\hat{\epsilon}\right)^{2} \rangle
\nonumber
\\
&=p_{0}R^{2}+p_{0}\left(\langle \hat{\epsilon}^{2}\rangle-\langle \hat{\epsilon}\rangle^{2}\right)  =\frac{\langle \hat{T}_{\phi \phi}\rangle}{\sin^{2}\theta}~,
\end{align}
where $\tilde{\sigma} \equiv \sigma/M$ and for simplicity hereafter we shall expand some expressions in powers of $\tilde{\sigma}\ll1$.
The components of the extrinsic curvature, as well as its trace, can be computed from the semiclassical junction condition in Eq.~(\ref{semi_class_junc}) on the three-dimensional $r=\textrm{constant} \equiv R$ hypersurface .
Finally, from Eq.~(\ref{semi_class_junc}) the background energy-density and pressure of the quantum fluid living on the membrane are obtained, i.e.,
\begin{align}
\rho_{0}&=-\frac{f(R)^{3/2}}{4\pi R}
\nonumber
\\
&\,\,\,\times\left[\frac{1}{f(R)+\frac{1}{2}f''(r_{+})\left(\langle \hat{\epsilon}^{2}\rangle-\langle \hat{\epsilon}\rangle^{2}\right)+\mathcal{O}(\tilde{\sigma}^{3})}\right]~,
\label{energydensity}
\\
\nonumber
\\
p_{0}&=\frac{R\left[2f(R)+Rf'(R) \right]}{16\pi\sqrt{f(R)} \left[R^{2}+\left(\langle \hat{\epsilon}^{2}\rangle-\langle \hat{\epsilon}\rangle^{2}\right)\right]}~.
\label{pressure}
\end{align}
In the above expressions, the quantum corrections arise through the $\langle \hat{\epsilon}^{2}\rangle-\langle \hat{\epsilon}\rangle^{2}$ term and also the $R=r_{+}+\langle \hat{\epsilon}\rangle$ term. Note that, in the classical limit when the quantum correction $\sigma$ vanishes, we have $R\rightarrow r_{+}$ and hence the energy density identically vanishes while the pressure diverges. This behaviour is expected in the BH case in order to sustain a fluid on the horizon~\cite{Thorne:1986iy}. 

Given the expressions for $\rho_{0}$ and $p_{0}$, one can expand the combination $(\rho_{0}+p_{0})$ as a power series in $\sigma$, and to the leading order we obtain
\begin{align} \label{rho0p0}
\rho_{0}+p_{0}&= \frac{f'(r_{+})}{16\pi \sqrt{f(r_{+}+\langle \hat{\epsilon} \rangle)}} + \mathcal{O}(\sigma) 
\nonumber
\\
&= \frac{\sqrt{f'(r_{+})}}{16\pi }\frac{\pi^{1/4}}{\sqrt{\sigma}}+\mathcal{O}(\sqrt{\sigma})~.
\end{align}
The above equation will have an application in the discussion of the boundary conditions associated with the gravitational perturbations.  

\subsection{Effect of the quantum membrane on the boundary condition for gravitational perturbations}

In this section, we discuss how the presence of a quantum membrane may affect the gravitational perturbations. For simplicity, we restrict our analysis to the case of axial perturbations, however a similar analysis can be carried out for the polar sector as well (see Appendix \ref{AppB} for further details). In the case of axial perturbations, the only non-vanishing components of the metric perturbations are $\delta g_{t\phi}$ and $\delta g_{r\phi}$, arising out of the Regge-Wheeler gauge condition~\cite{Regge:1957td,Maggio:2020jml}. Owing to the spherical symmetry of the background spacetime, the two metric perturbations can be expressed in terms of Legendre polynomials as
\begin{align}
\delta g_{t \phi} &= e^{-i \omega t} h_0(r) \sin \theta \partial_\theta P_\ell(\cos \theta)~,\\
\delta g_{r \phi} &= e^{-i \omega t} h_1(r) \sin \theta \partial_\theta P_\ell(\cos \theta)~,
\end{align}
for a mode with azimuthal number $\ell \geq 2$.
The metric perturbations are related to the properties of the quantum membrane through the perturbation of the semiclassical junction conditions in Eq.~(\ref{semi_class_junc}). In particular, only two components of the extrinsic curvature are perturbed as
\begin{align}
\delta K_{t\phi}&=\frac{1}{2}e^{-i\omega t}\sqrt{f(R)}
\nonumber
\\
&\times\Big[i\omega h_{1}(R)+h_{0}'(R)\Big]\sin \theta \partial_{\theta}P_{\ell}(\cos\theta)~,
\label{K_tphi}
\\
\delta K_{\theta \phi}&=-\frac{1}{2}e^{-i\omega t}\sqrt{f(R)}h_{1}(R)
\nonumber
\\
&\times \left(-\cos\theta \partial_{\theta}+ \sin \theta \partial_{\theta}^{2}\right)P_{\ell}(\cos\theta)~.
\label{K_thetaphi}
\end{align}
As a consequence, the perturbation of the trace of the extrinsic curvature is $\delta K=K_{ab}\delta h^{ab}+h^{ab}\delta K_{ab}$. Since $h^{t\phi}=0=h^{\theta \phi}$ and $K_{t\phi}=0=K_{\theta \phi}$, it follows that the perturbation of the trace of the extrinsic curvature identically vanishes for axial perturbations, i.e., $\delta K=0$. 

The junction conditions also involve the stress-energy tensor of the quantum membrane whose perturbations must be taken into account. 
An explicit computation yields the following components for the expectation values of the perturbed stress-energy tensor:
\begin{align}
\langle\delta \hat{T}_{t\phi}\rangle&=-e^{-i\omega t}\rho_{0}(R)h_{0}(R)\sin \theta \partial_{\theta}P_{\ell}(\cos\theta)
\nonumber 
\\
&-\sin^{2}\theta \delta u^{\phi}\left(\rho_{0}+p_{0}\right)\langle \hat{R}^{2}\sqrt{f(\hat{R})}\rangle~,
\label{pert_tphi}
\\
\langle\delta \hat{T}_{\theta \phi}\rangle&=-\eta \sin^{2}\theta \left(\partial_{\theta}\delta u^{\phi}\right)\langle \hat{R}^{2}\rangle
\nonumber
\\
&=-\eta \sin^{2}\theta \partial_{\theta}\delta u^{\phi}\left[R^{2}+\left(\langle \hat{\epsilon}^{2}\rangle-\langle \hat{\epsilon}\rangle^{2}\right)\right]~.
\label{pert_thetaphi}
\end{align}
Here $\delta u^{\phi}$ denotes the perturbation in the velocity of the membrane fluid due to the gravitational perturbations. Note that, for axial gravitational perturbations, only the $\phi$-component of the three-velocity is perturbed, while the rest of the components retain their background values. 
The computation of the expectation value in Eq.~(\ref{pert_tphi}) requires a careful analysis.
Indeed, due to the presence of the square root, the above quantity is nonanalytic in the operator $\hat{\epsilon}$ and hence the computation of the expectation value has ambiguities. Thus, we replace the above expectation value by the following one,
\begin{align}
\langle \hat{R}^{2}\sqrt{f(\langle\hat{R}\rangle)}~\rangle &=\sqrt{f(R)}\langle r_{0}^{2}+2r_{0}\hat{\epsilon}+\hat{\epsilon}^{2} \rangle
\nonumber
\\
&=R^{2}\sqrt{f(R)}+\sqrt{f(R)}\left(\langle \hat{\epsilon}^{2}\rangle-\langle \hat{\epsilon}\rangle^{2} \right)~,
\end{align}
where the location of the quantum membrane and the quantum correction are given in Eq.~(\ref{R}) and Eq.~(\ref{qdiff_caseII}), respectively. Substituting the above result in Eq.~(\ref{pert_tphi}), we obtain the desired expectation value for the $(t,\phi)$ component of the perturbed stress-energy tensor. 

In general, the perturbation of the geometrical term in the left hand side of Eq.~(\ref{semi_class_junc}) yields $\delta K_{ab}-K\delta h_{ab}-h_{ab}\delta K$. Since $\delta K$ identically vanishes, the perturbed semiclassical junction condition for axial gravitational perturbations reads
\begin{align}
\delta K_{ab}-K\delta h_{ab}=-8\pi\langle \delta \hat{T}_{ab}\rangle~,
\end{align}
where the only non-vanishing components are the $(t,\phi)$ and the $(\theta,\phi)$ terms. 
The $(t,\phi)$ component of the above equation together with Eq.~(\ref{K_tphi}) and Eq.~(\ref{pert_tphi})
can be used to derive $\delta u^{\phi}$, which takes the form
\begin{align}\label{delta_uphi}
\delta u^{\phi}&=\frac{e^{-i\omega t}\partial_{\theta}P_{\ell}(\cos\theta)}{8\pi \sin\theta \left(\rho_{0}+p_{0}\right)\sqrt{f(R)}\left[R^{2}+\left(\langle \hat{\epsilon}^{2}\rangle-\langle \hat{\epsilon}\rangle^{2} \right)\right]} \nonumber
\\
&\times \Bigg[\frac{1}{2}\sqrt{f(R)}\left[i\omega h_{1}(R)+h_{0}'(R)\right]
 \nonumber
\\
&- \left(\frac{f'(R)}{2\sqrt{f(R)}}+\frac{2\sqrt{f(R)}}{R}\right)h_{0}(R)-8\pi \rho_{0}(R)h_{0}(R)\Bigg] \,.
\end{align}
Similarly, using the expression for $\delta K_{\theta \phi}$ from Eq.~(\ref{K_thetaphi}) and the expectation value of $\delta \hat{T}_{\theta \phi}$ from Eq.~(\ref{pert_thetaphi}), the $(\theta,\phi)$ component of the perturbed semiclassical junction condition can be obtained. Then, substitution of $\delta u^{\phi}$ from Eq.~(\ref{delta_uphi}) in the $(\theta,\phi)$ component of the junction condition yields the following expression for $h_{1}(R)$, i.e., the radial part of the $(r,\phi)$ component of metric perturbation on the classical location of the membrane,
\begin{align}\label{h1R}
h_{1}(R)&= \frac{-2\eta}{\left(\rho_{0}+p_{0}\right)f(R)} \Bigg[\frac{1}{2}\sqrt{f(R)}\left[i\omega h_{1}(R)+h_{0}'(R)\right]
\nonumber
\\
&- \left(\frac{f'(R)}{2\sqrt{f(R)}}+\frac{2\sqrt{f(R)}}{R}\right)h_{0}(R) - 8\pi \rho_{0}(R)h_{0}(R)\Bigg]~.
\end{align}
The radial part of the remaining metric perturbation, namely $h_{0}(r)$, can be determined in terms of $h_{1}(r)$ through the relation $h_{0}(r)=-f(r)\partial_r[f(r)h_{1}(r)]/(i\omega)$~\cite{Regge:1957td}. 
Finally, it is convenient to impose the boundary condition on the Regge-Wheeler function defined as $\psi(r)\equiv f(r)h_{1}(r)/r$. Replacing $h_{1}(r)$ and $h_{0}(r)$ from Eq.~(\ref{h1R}) in favour of the Regge-Wheeler function and its derivative with respect to the tortoise coordinate $x$ (defined through $dr/dx=f(r)$), we obtain the following boundary condition:
\begin{widetext}
\begin{align}\label{boundary}
i\omega \psi(R)&=\frac{\eta}{\left(\rho_{0}+p_{0}\right)\sqrt{f(R)}}\Bigg[V_{\rm axial}(R)\psi(R)
-\frac{1}{R}\frac{d\psi(R)}{dx}\left[Rf'(R)-2f(R)\right]
\nonumber
\\
&-\frac{4f(R)}{R}\left(\frac{d\psi(R)}{dx}+\frac{f(R)}{R}\psi(R)\right)\left(1+\frac{4\pi \rho_{0}R}{\sqrt{f(R)}}\right)\Bigg]~.
\end{align}
\end{widetext}
Here we have used the differential equation satisfied by the Regge-Wheeler function, which reads~\cite{Regge:1957td}
\begin{align}\label{ODE}
\frac{{\rm d}^2\psi}{{\rm d}x^2}+\left[\omega^2-V_{\rm axial}(r)\right]\psi=0 \,,
\end{align}
where $V_{\rm axial}(r)$ is the potential associated with the axial gravitational perturbations,
\begin{align}
V_{\rm axial}(r)=f(r)\left[\frac{\ell(\ell+1)}{r^{2}}-\frac{f'(r)}{r}-2\left(\frac{1-f(r)}{r^{2}}\right)\right]~.
\end{align}
For the Schwarzschild metric, the axial potential reduces to $V_{\rm axial}=(1-2M/r)[\ell(\ell+1)/r^{2}-6M/r^{3}]$.

In the limit of vanishing quantum corrections, from Eq.~(\ref{energydensity}) it follows that $\rho_{0}=-\sqrt{f(R)}/(4\pi R)$, and hence the last term in Eq.~(\ref{boundary}) identically vanishes. On the other hand, the presence of $\sqrt{f(R)}$ in the denominator of Eq.~(\ref{boundary}) appears problematic, since $f(R)$ vanishes in the limit of vanishing quantum correction, leading to a divergent contribution. However, the term $(\rho_{0}+p_{0})$ appearing in the denominator of the boundary condition diverges as the membrane becomes classical, such that the combination $\sqrt{f(R)}(\rho_{0}+p_{0})\sim f'(r_{+})/(16\pi)$ is finite. Therefore, the boundary condition remains finite in the limit of a vanishing quantum correction, namely
\begin{align} \label{BHbc}
i\omega\psi(R)=-16\pi \eta\frac{d\psi(R)}{dx}~,
\end{align}
which, for $\eta=\eta_{\rm BH}$, coincides with the appropriate boundary condition on a classical BH horizon.  This is expected, since in the limit of vanishing quantum correction, the surface of the membrane returns back to $r_{+}$ and the physical system becomes identical to that of a classical BH. 

Let us focus on the corrections introduced by the quantum membrane. The boundary condition can be computed analytically for generic values of $\tilde{\sigma}$ but the final expression is cumbersome. Thus, we explicitly provide its expression in the limit $\tilde{\sigma}\ll 1$, which is anyway expected from microscopical corrections of an astrophysical BH. To ${\cal O}(\tilde{\sigma}^3)$ corrections, the boundary condition reads
\begin{widetext}
\begin{align}
\left[i\omega - \frac{4\left(\ell (\ell+1) - 3\right) \sqrt{\pi} \eta \tilde{\sigma}}{M} - \frac{2 \ell (\ell + 1) \eta \tilde{\sigma}^2}{M}  \right] \psi(R)=- 16 \pi \eta \left[ 1 + \frac{(\pi - 2) \tilde{\sigma}^2}{8 \pi} \right] \frac{d\psi(R)}{dx}~.
\end{align}
\end{widetext}
Notice that the boundary condition above modifies the BH boundary condition in Eq.~(\ref{BHbc}) due to the quantum nature of the membrane.
As a consequence, it is expected that the reflective properties of the object and its QNM spectrum are affected by the quantum corrections. We shall investigate these effects in the next sections.

\subsection{Reflectivity of the quantum membrane} \label{sec:reflectivity}

The modified boundary condition in Eq.~(\ref{boundary}) affects the reflectivity of the membrane. The latter can be computed analytically when the effective potential at the inner boundary is negligible. Imposing $\omega^2\gg V_{\rm axial}(R)$ implies 
\begin{align}
    \tilde{\sigma} \ll \left(\frac{8 \sqrt{\pi}}{\ell(\ell+1) - 3}\right)\omega^2M^2 \,. \label{condsigma}
\end{align}
In this case, the generic solution of Eq.~(\ref{ODE}) near the membrane reads 
\begin{align} \label{asymptrefl}
    \psi_{\rm M} = e^{-i\omega x} + \mathcal{R}e^{i\omega x} \,, \qquad x \to x(R) \,,
\end{align}
where $\mathcal{R}$ is the reflectivity of the membrane.
From Eq.~(\ref{asymptrefl}), it follows that
\begin{align}\label{def_reflectivity}
\mathcal{R}=\left(\frac{\omega \psi_{\rm M}-i d\psi_{\rm M}/dx}{\omega \psi_{\rm M}+i d\psi_{\rm M}/dx}\right)e^{-2i\omega x}~,
\end{align}
where the expression for $(d\psi/dx)/\psi$ on the surface $r=R$ can be derived from the boundary condition in Eq.~(\ref{boundary}). 
When this expression is substituted in Eq.~(\ref{def_reflectivity}), it
yields the desired analytical expression for the reflectivity. For this purpose, we rewrite the boundary condition presented in Eq.~(\ref{boundary}) in the following manner,
\begin{widetext}
\begin{align}\label{boundary_modified}
i\omega \psi(R)&=\frac{\eta}{\left(\rho_{0}+p_{0}\right)\sqrt{f(R)}}\Bigg[\frac{d\psi(R)}{dx}\left(-\frac{Rf'(R)+2f(R)}{R}-16\pi \rho_{0} \sqrt{f(R)}  \right)
\nonumber
\\
&\hskip 3 cm -4\left(\frac{f(R)}{R}\right)^{2}\left(1+\frac{4\pi \rho_{0}R}{\sqrt{f(R)}}\right)\psi(R)\Bigg]
\equiv-B\frac{d\psi(R)}{dx}-A\psi(R)~,
\end{align}
\end{widetext}
where, in the last line, we have introduced the quantities 
\begin{align}
A&\equiv \frac{\eta}{4\pi \eta_{\rm BH}}\frac{f(R)^{3/2}}{\left(\rho_{0}+p_{0}\right)R^{2}}\left(1+\frac{4\pi \rho_{0}R}{\sqrt{f(R)}}\right)~,
\label{def_A}
\\
B&\equiv \frac{\eta}{\eta_{\rm BH}}\left[\frac{\rho_{0}}{\rho_{0}+p_{0}}+\frac{1}{16\pi(\rho_{0}+p_{0})}\frac{2f(R)+Rf'(R)}{R\sqrt{f(R)}}\right]~.
\label{def_B}
\end{align}
\begin{figure*}[t!]
\centering
\includegraphics[width=0.48\textwidth]{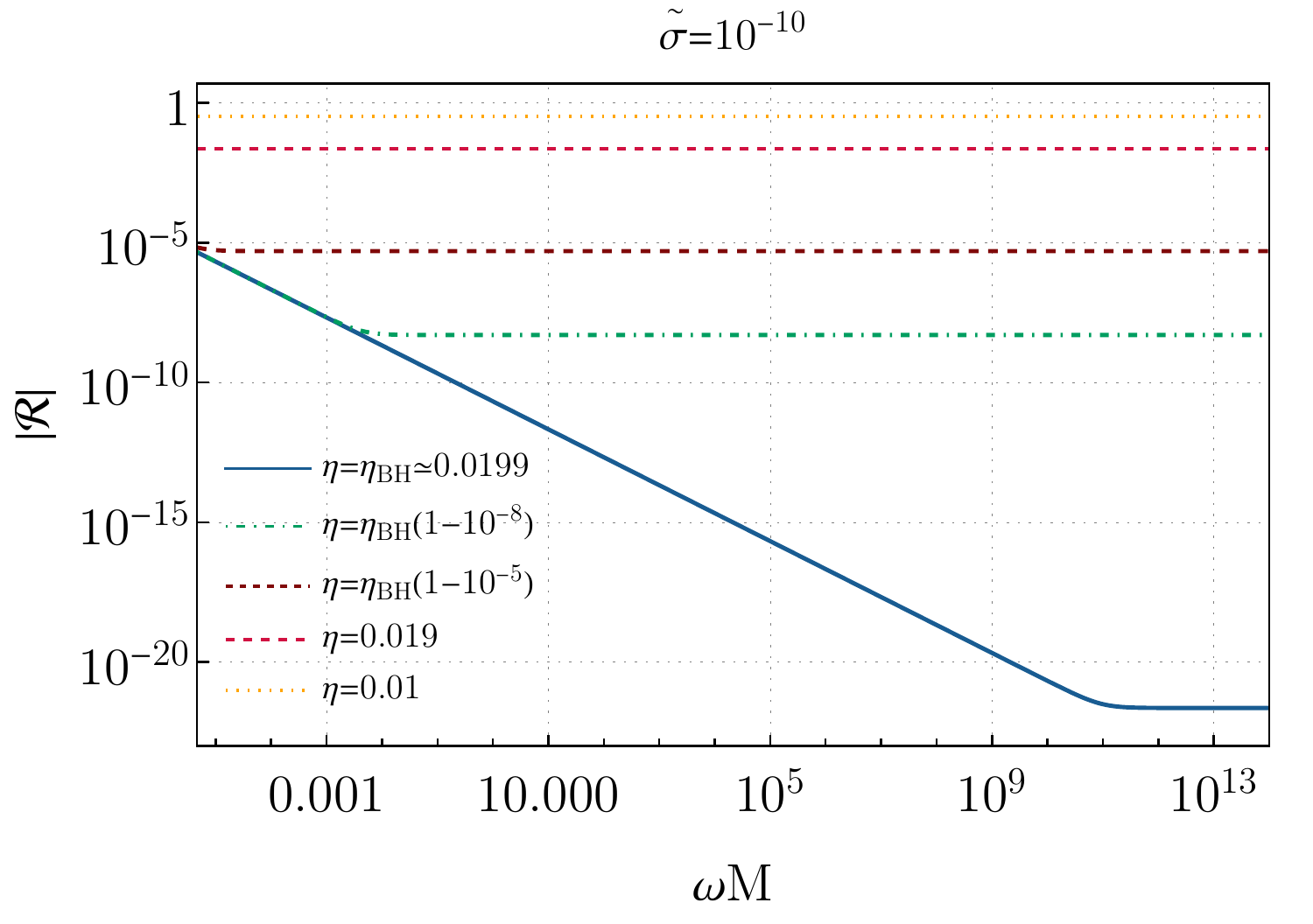}
\includegraphics[width=0.48\textwidth]{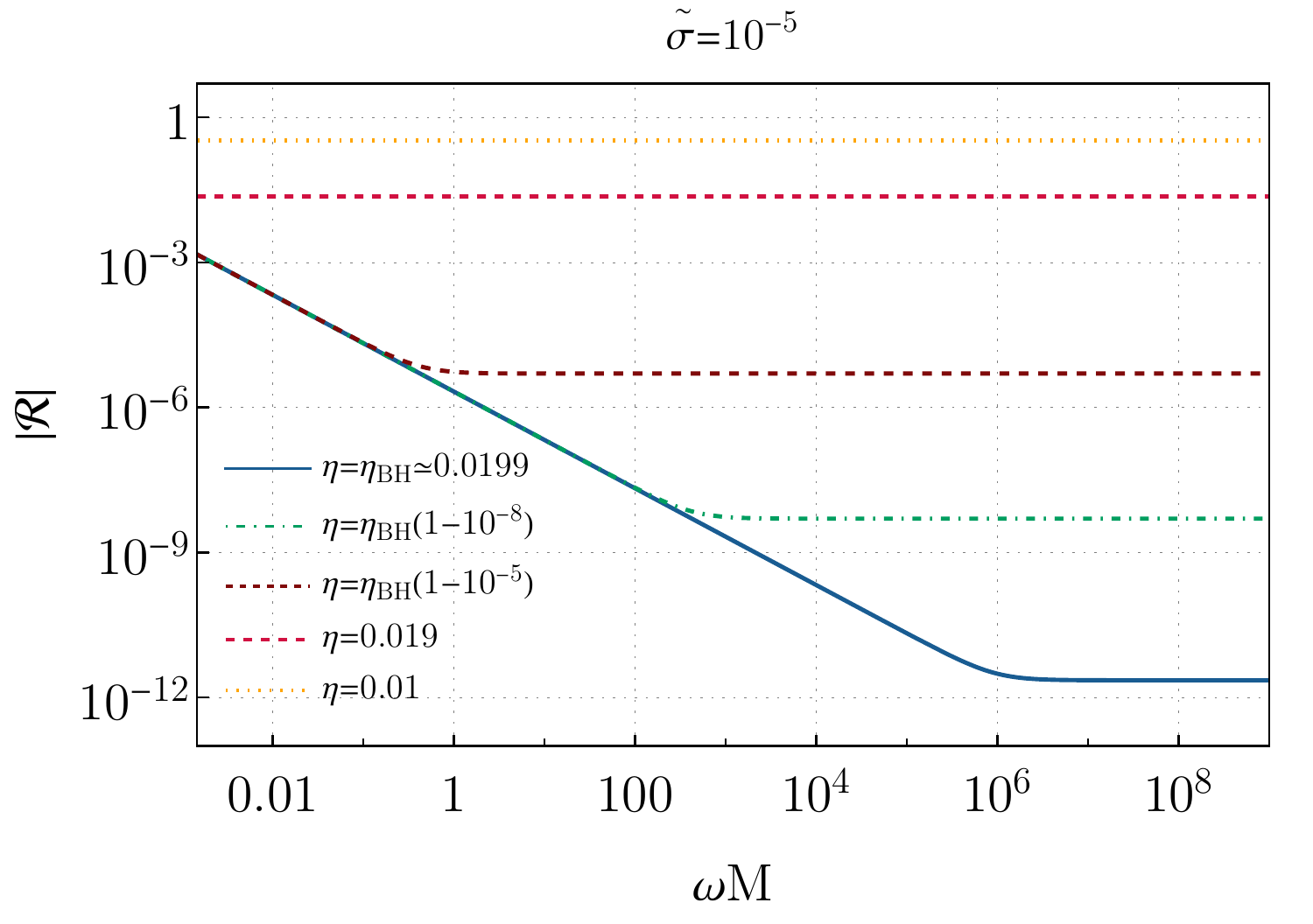}
\caption{The reflectivity of the quantum membrane as a function of the frequency for $\tilde{\sigma}=10^{-10}$ (left panel) and $\tilde{\sigma}=10^{-5}$ (right panel) for various choices of the membrane viscosity $\eta$ and $\ell=2$. We only consider frequencies for which Eq.~(\ref{condsigma}) is satisfied.
}
\label{fig:reflectivitysigma}
\end{figure*}
Finally, from Eq.~(\ref{def_reflectivity}) we get
\begin{align}
\mathcal{R}=\frac{\omega(B-1)+iA}{\omega(B+1)-iA}~\,,
\end{align}
which, as we remind, is valid when the condition in Eq.~(\ref{condsigma}) is satisfied.
Note that the reflectivity depends on the ratio $\eta/\eta_{\rm BH}$, as well on the quantum parameter $\tilde{\sigma}$ and the frequency. Up to ${\cal O}(\tilde{\sigma}^3)$, the square of the magnitude of the reflectivity reads
\begin{align}
|\mathcal{R}|^2 \sim \left(\frac{1-\eta / \eta_{\rm BH}}{1+\eta / \eta_{\rm BH}}\right)^2 + \frac{16384 \left[\ell (\ell+1) -3\right]^2 \pi^{3} \eta^4 \tilde{\sigma}^2}{\left(1+\eta/\eta_{\rm BH}\right)
^4 \omega^2 M^2}~.
\end{align}
The frequency dependence of the reflectivity in this regime is shown in Fig.~\ref{fig:reflectivitysigma}. Interestingly, even when $\eta=\eta_{\rm BH}$, the reflectivity is nonzero due to the quantum properties of the membrane, i.e., $|{\cal R}|^2\sim \tilde{\sigma}^2/(\omega^2M^2)$. Furthermore, in the large frequency limit and for $\eta=\eta_{\rm BH}$,
\begin{widetext}
\begin{align}
    |{\cal R}(\omega\to\infty)| &\sim 2 (2-\pi) \pi  \tilde\sigma ^2 \left(-8 \pi ^{3/2}+2 (\pi-6) \pi  \tilde\sigma +7 (\pi-2)
   \sqrt{\pi } \tilde\sigma ^2+3 (\pi-2) \tilde\sigma ^3\right) \times \nonumber \\
   &\left(
    256 \pi ^{7/2}-128 (\pi-2) \pi ^3
   \tilde\sigma +16 (\pi-14 ) (\pi-1) \pi ^{5/2} \tilde\sigma ^2+4 \pi ^2 (100+\pi  (11 \pi-80 ))
   \tilde\sigma ^3\right.\nonumber\\
   &\left. +2 (\pi-2) \pi ^{3/2} (29 \pi -66) \tilde\sigma ^4 +34 (\pi-2)^2 \pi  \tilde\sigma ^5+9
   (\pi-2)^2 \sqrt{\pi } \tilde\sigma ^6+(\pi-2)^2 \tilde\sigma ^7\right)^{-1}\,,
\end{align}
\end{widetext}
which is independent on $\ell$ and, to leading order in $\tilde\sigma\ll1$, reduces to
\begin{align}
    |\mathcal{R}(\omega\to\infty)| \sim \frac{(\pi-2)\tilde{\sigma}^2}{16 \pi}  \,.
\end{align}
In the $\tilde{\sigma} \rightarrow 0$ limit, the reflectivity identically vanishes, since for vanishing quantum corrections and $\eta=\eta_{\rm{BH}}$ the membrane mimics the properties of a classical BH. However, for $\tilde{\sigma}\neq0$, the reflectivity is constant in the large-frequency limit, which explains the plateau of the blue curve ($\eta=\eta_{\rm BH}$) in Fig.~\ref{fig:reflectivitysigma}.

\subsection{QNM spectrum and ringdown for the quantum membrane}

After imposing outgoing boundary conditions at infinity and Eq.~(\ref{boundary}) at the location of the quantum membrane, Eq.~(\ref{ODE}) defines an eigenvalue problem whose complex eigenvalues are the QNMs of the object, $\omega=\omega_{\rm R}+i\omega_{\rm I}$. Here, $\omega_{\rm R}$ is the real part of the QNMs, denoting the oscillatory nature of the eigenstate, while $\omega_{\rm I}$ is the imaginary part of the QNMs, denoting its characteristic exponential decay. In our convention, a stable mode has $\omega_{\rm I}<0$, whereas an unstable mode has $\omega_{\rm I}>0$. 

\begin{figure*}[th]
\centering
\includegraphics[width=0.49\textwidth]{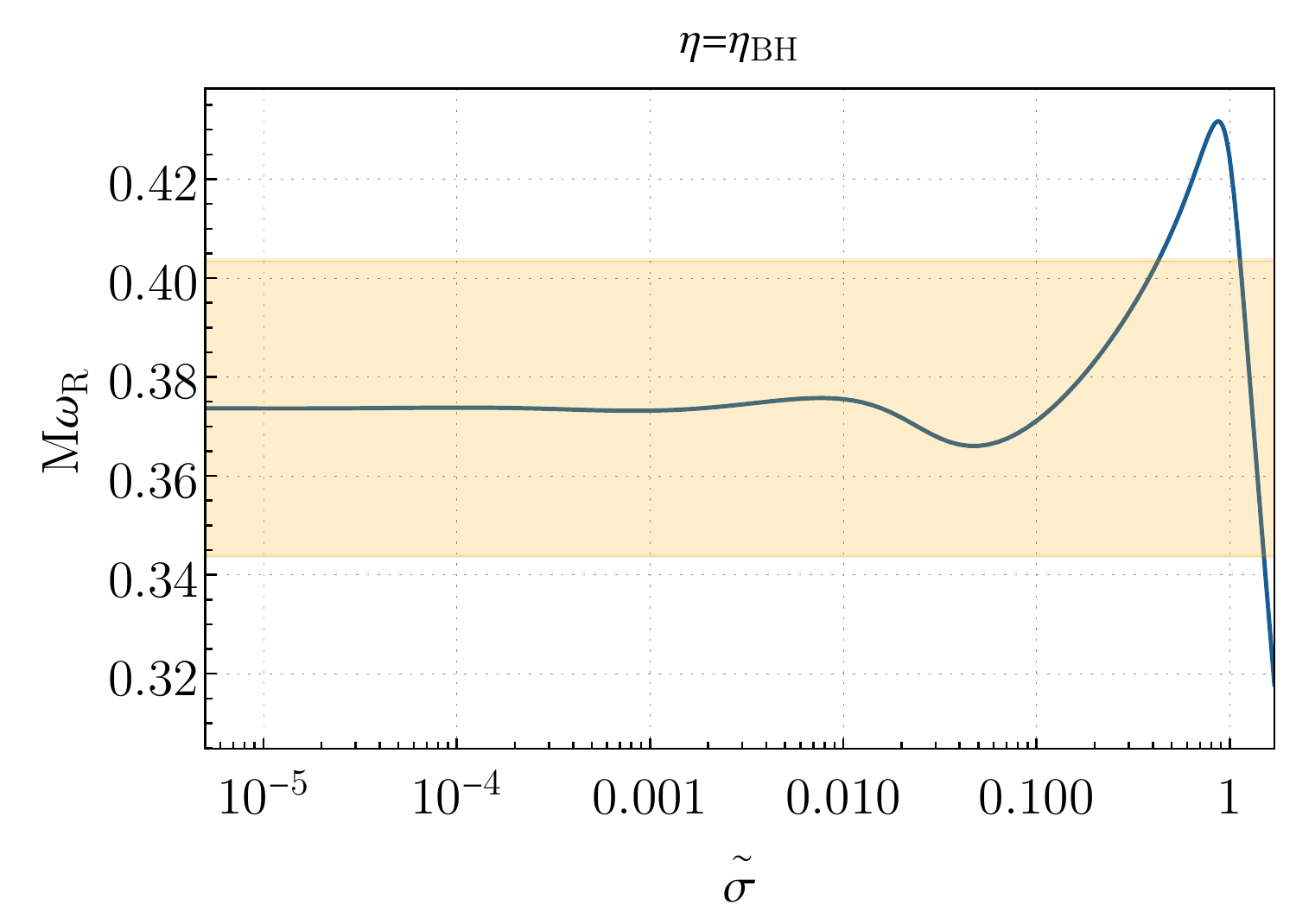}
\includegraphics[width=0.49\textwidth]{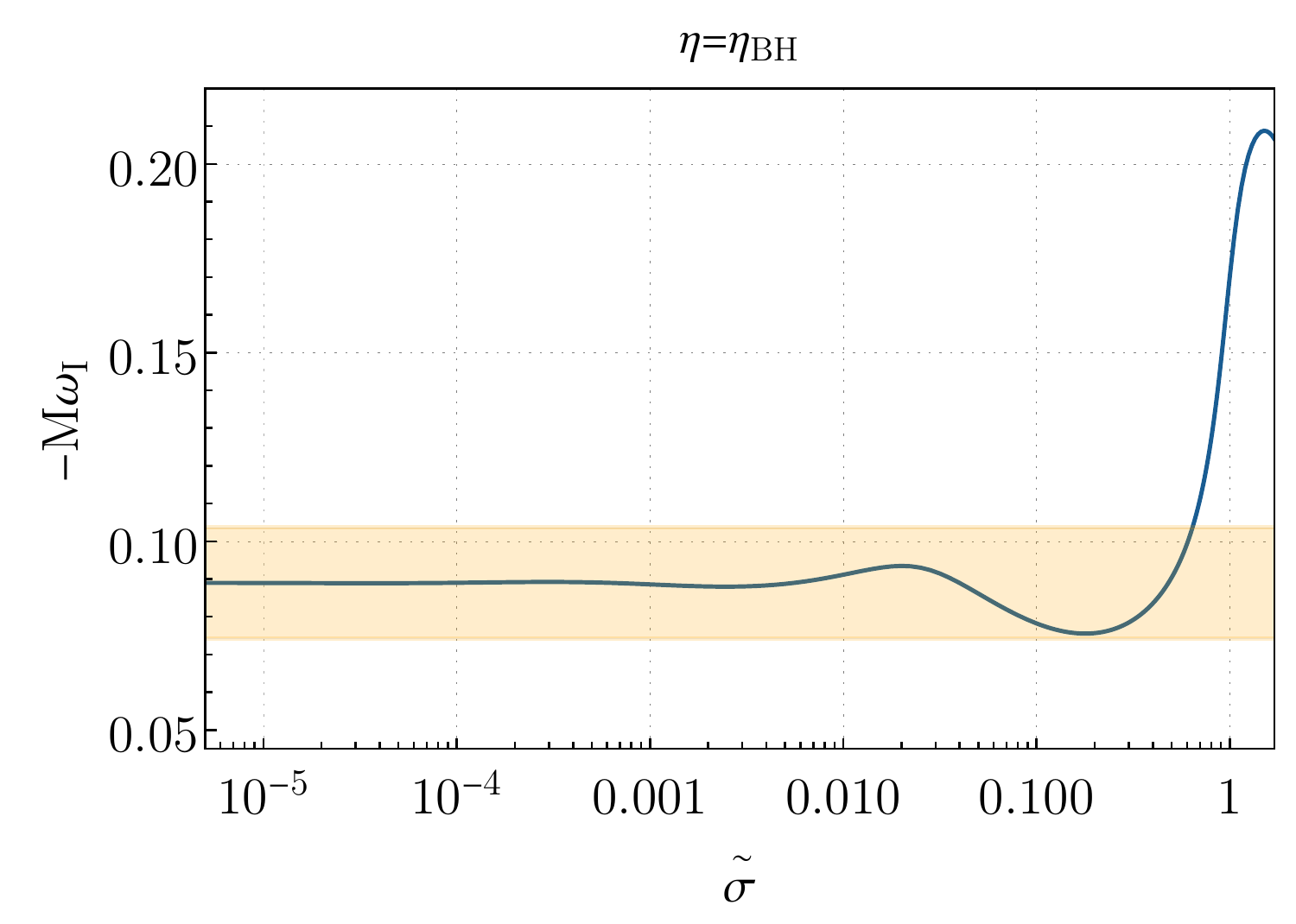}
\caption{Real (left panel) and imaginary (right panel) part of the fundamental $\ell=2$ QNM of a quantum membrane with effective shear viscosity $\eta=\eta_{\rm BH}$ as a function of the dimensionless parameter $\tilde{\sigma}$ which is related to the quantum nature of the membrane. The classical BH limit is recovered when $\tilde{\sigma}\to0$.
The highlighted regions are the maximum allowed deviations for the least damped QNM of GW150914~\cite{TheLIGOScientific:2016src,Ghosh:2021mrv}. 
Quantum deviations would be measurable by current GW detectors when $\tilde{\sigma} \gtrsim 0.5$.
}
\label{fig:QNMsquantummembrane}
\end{figure*}

Given the boundary conditions, we compute the QNM spectrum using the continued fractions method in a variant adapted from the case of compact stars~\cite{Pani:2009ss,Maggio:2020jml}. The result of this analysis is presented in Fig.~\ref{fig:QNMsquantummembrane}, where the real (left panel) and imaginary (right panel) parts of the fundamental $\ell=2$ QNMs of the quantum membrane are shown as a function of the quantum correction $\tilde{\sigma}$, with $\eta=\eta_{\rm BH}$. The highlighted regions in Fig.~\ref{fig:QNMsquantummembrane} correspond to the maximum allowed deviations (with $90\%$ credibility) for the least-damped QNM in the event GW150914~\cite{TheLIGOScientific:2016src,Ghosh:2021mrv} with respect to the BH case, corresponding to a range $\sim 16\%$ and $\sim 33\%$ for the real and the imaginary part of the QNM, respectively. 
Fig.~\ref{fig:QNMsquantummembrane} shows that the deviations from the BH QNM due to quantum corrections would be measurable by current GW detectors only when $\tilde{\sigma} \gtrsim 0.5$. It is worth highlighting that the structure of the QNM spectrum is similar to the case of a  classical membrane discussed in~\cite{Maggio:2020jml}. 
In that case, the membrane is located at $R=2M(1+\delta)$, where $\delta$ is a classical displacement. The relation among the locations of the classical and the quantum membrane is $\delta = \tilde{\sigma}/(2\sqrt{\pi})$, but the boundary conditions are not the same.
Indeed, the two scenarios are conceptually very different. In the classical case, the separation of the membrane from the horizon is introduced in an ad-hoc manner, while in the present context the separation from the classical horizon arises naturally from the quantum properties of the membrane.

The fundamental QNM provides only partial information on the linear response of the object. Indeed, when $\tilde{\sigma} \ll 1$, the classical radius of the object is $R\sim r_+$ and the perturbations take a long time before probing the inner boundary. 
This automatically results in a prompt ringdown which is nearly identical to the BH one and can differ from it at later times due to the nontrivial reflectivity of the quantum membrane~\cite{Cardoso:2016rao,Cardoso:2016oxy,Cardoso:2017cfl,Maggio:2020jml}.

\begin{figure*}[th]
\centering
\includegraphics[width=0.49\textwidth]{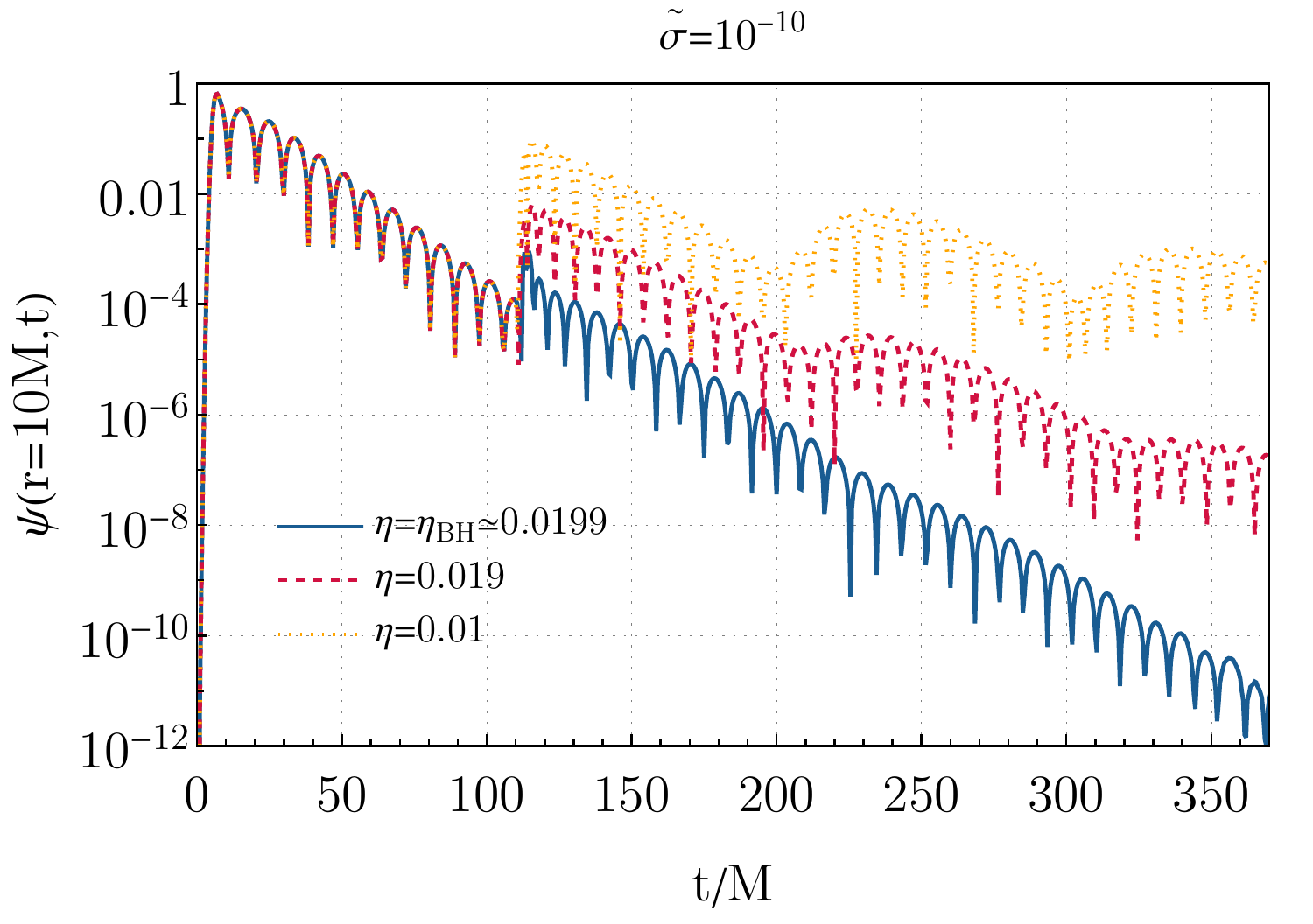}
\includegraphics[width=0.49\textwidth]{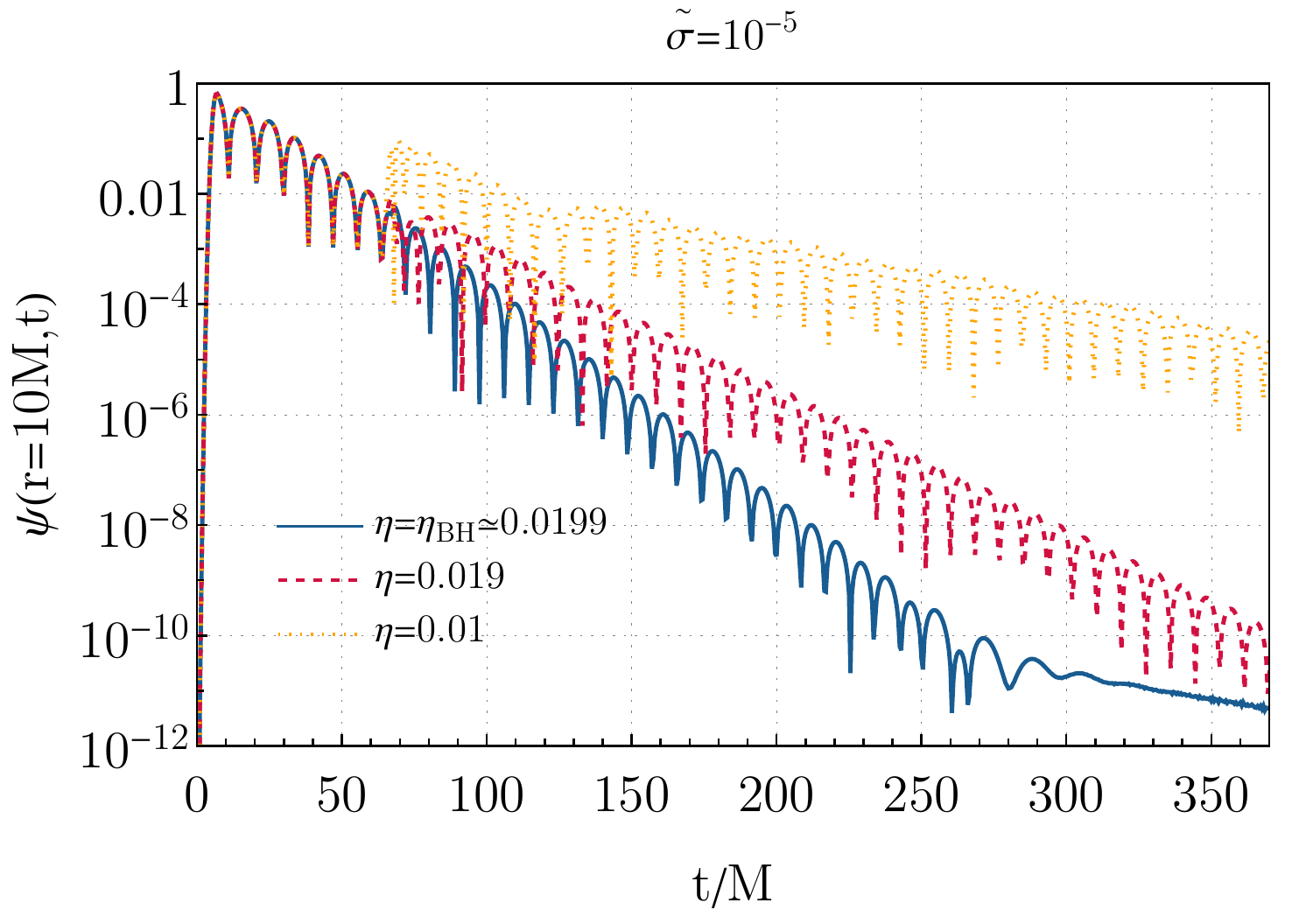}
\caption{Ringdown waveform as a function of time for $\tilde{\sigma}=10^{-10}$ (left panel) and $\tilde{\sigma}=10^{-5}$ (right panel) for several choices of the shear viscosity of the quantum membrane. Due to the nonvanishing reflectivity of the membrane, the GW signal displays echoes after regular intervals of time, even when $\eta=\eta_{\rm BH}$. The initial perturbation is a Gaussian wavepacket where $\psi(x,0)=0$ and $\partial_t \psi(x,0) = \exp[-(x-7)^2]$.
}
\label{fig:waveform}
\end{figure*}

In order to better understand this behavior, let us look at the time-domain response of the system. This is presented in Fig.~\ref{fig:waveform} for $\tilde{\sigma}=10^{-10}$ (left panel) and $\tilde{\sigma}=10^{-5}$ (right panel), and for different choices of the shear viscosity of the quantum membrane.
The initial perturbation has a Gaussian profile where $\partial_t \psi(x,0) = \exp[-(x-7)^2]$ and $\psi(x,0)=0$.
Due to the presence of the reflective membrane arising out of quantum effects, the waveform shows the recursive appearance of GW echoes. This is most evident for $\eta=0.01$, for which the reflectivity of the membrane is almost unity, as shown in Fig.~\ref{fig:reflectivitysigma}.
Interestingly, even for $\eta=\eta_{\rm BH}$ the postmerger signal is characterized by the appearance of the first GW echo. However, the amplitude of the first echo is much smaller than the prompt ringdown signal, and subsequent GW echoes are highly suppressed. As a consequence,  detecting GW echoes from a quantum horizon-like membrane  would be challenging with current GW detectors.

Note that the time delay between subsequent echoes is consistent with the roundtrip time that the radiation takes to probe the boundary,
\begin{align} \label{tauecho}
    \Delta t = 2 M \left[1 - \frac{\tilde{\sigma}}{\sqrt{\pi}} -2 \log \left(\frac{\tilde{\sigma}}{\sqrt{\pi}} \right) \right] \,.
\end{align}
For the parameters considered in the left (right) panel of Fig.~\ref{fig:waveform}, $\Delta t\simeq 100M$ ($\Delta t\simeq 50M$) for any $\eta$. 

The amplitude of the GW echoes with respect to the prompt ringdown depends on the reflectivity of the quantum membrane and the transmission coefficient of the photon sphere.
It is important to stress that the reflectivity computed in Sec.~\ref{sec:reflectivity} is valid only when Eq.~(\ref{condsigma}) is satisfied. This condition might be violated for certain frequency components of the initial wavepacket. This might explain some interesting effects such as a sizeable amplification of the wavepacket for $\eta=\eta_{\rm BH}$ when the reflectivity is anyway small.
Although such amplification is small (note that the vertical scale of Fig.~\ref{fig:waveform} is logarithmic), this feature provides an important generic signature that might be looked for in high-precision data.

\begin{figure}[th]
\centering
\includegraphics[width=0.49\textwidth]{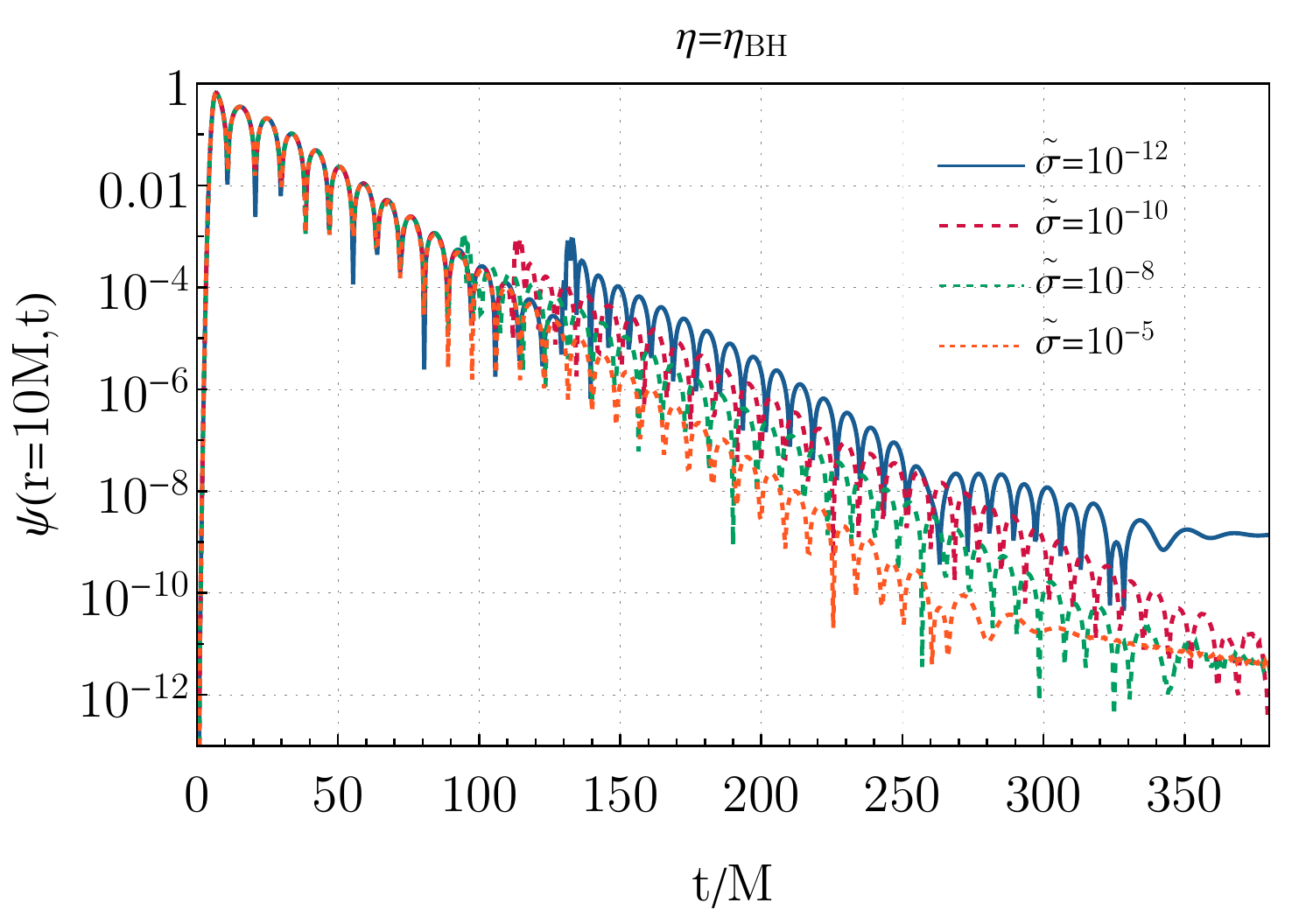}
\caption{Ringdown waveforms as a function of time for different choices of the quantum correction $\tilde{\sigma}$ and for a quantum membrane with shear viscosity  $\eta=\eta_{\rm BH}$. The figure shows that the quantum parameter $\tilde{\sigma}$ can have observable effects on the ringdown signal.}
\label{fig:waveformetabh}
\end{figure}

Let us focus on the scenario with $\eta=\eta_{\rm BH}$. Fig.~\ref{fig:waveformetabh} shows the time-domain ringdown waveforms (in a logarithmic scale) for several locations of the quantum membrane. As the quantum correction decreases, the time delay of the first GW echo is longer due to the logarithmic dependence in Eq.~(\ref{tauecho}).
This makes the echoes more evident on the logarithmic scale used in Fig.~\ref{fig:waveformetabh}, since the reflected echo signal stands out the prompt ringdown, which has already been exponentially damped by the time the first echo arrives.
This also explains why in Fig.~\ref{fig:waveform} the echoes are apparently less noticeable for $\tilde{\sigma}=10^{-5}$ rather than for $\tilde{\sigma}=10^{-10}$. This is not because of the reduced reflectivity, but because of the shorter echo delay time, which results into the echoes to fall within the domain of the primary signal.\footnote{
From Fig.~\ref{fig:reflectivitysigma} one notes that for $\eta\neq \eta_{\rm BH}$, the reflectivity is almost constant, however the echo time delay changes significantly with $\tilde{\sigma}$, see Eq.~(\ref{tauecho}).}
Note however that the echo morphology is complex due to various effects: (i) when $\tilde\sigma$ increases, the effective size of the cavity decreases and therefore long-lived modes are less efficiently trapped; (ii) the effective reflectivity of the membrane at the relevant frequencies (not necessarily within the range given by Eq.~(\ref{condsigma}) in which the reflectivity can be computed easily) depends generically on $\tilde\sigma$ and $\omega$ in a nontrivial way.

Another interesting feature is that, for certain values of $\tilde{\sigma}$, the waveform is approximately constant at late times (see, e.g., the blue curve in Fig.~\ref{fig:waveformetabh}). Although we do not have a clear explanation for this behavior, we suspect it might be due to the peculiar boundary conditions that, in the time domain, have the schematic form
\begin{equation}
    \partial_x \psi(x,t) = \alpha \partial_t \psi(x,t)+\beta \psi(x,t)\,,
\end{equation}
where $\alpha$ and $\beta$ are two coefficients depending on $\eta$ and $\tilde{\sigma}$. In the $\tilde{\sigma}\to0$ limit, $\alpha\to 1$ and $\beta\to0$ but their structure can affect the signal at finite values, e.g. by introducing a sort of ``memory effect'' at the linear level (see, e.g., the recent discussion in~\cite{Garfinkle:2022dnm}) as shown in Fig.~\ref{fig:waveformetabh}. This effect is interesting on its own and will be investigated in detail elsewhere.

\begin{figure}[th]
\centering
\includegraphics[width=0.49\textwidth]{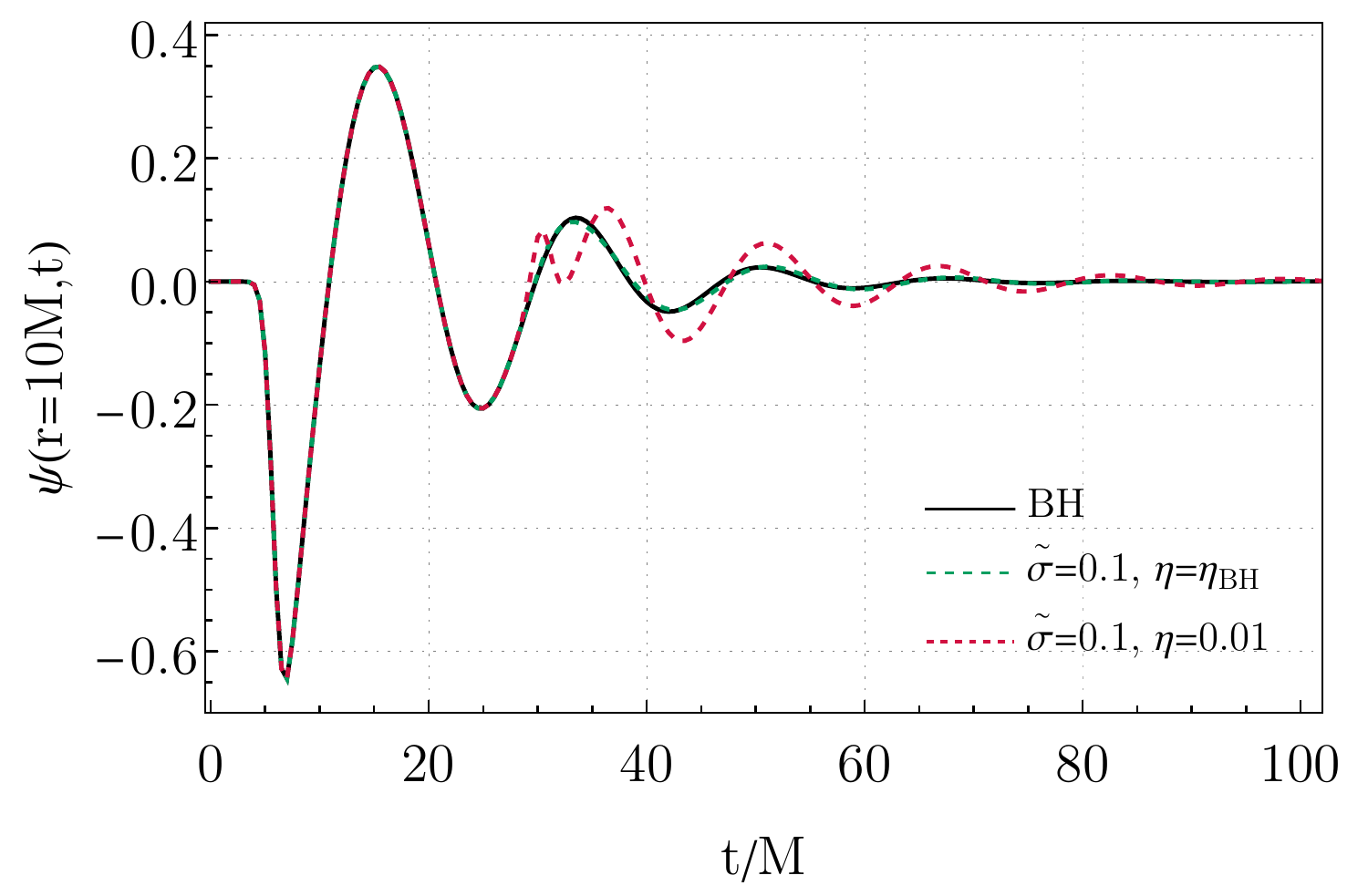}
\caption{Ringdown waveform as a function of time in a linear scale for $\tilde{\sigma}=0.1$, with two different choices of the shear viscosity  of the quantum membrane and compared to the classical BH case. The $\eta=\eta_{\rm BH}$ case is practically indistinguishable from the classical BH case on the linear scale of the plot.
}
\label{fig:waveformlargesigma}
\end{figure}

Finally, Fig.~\ref{fig:waveformlargesigma} shows the ringdown waveform for a 
relatively large value of the quantum correction, i.e. $\tilde{\sigma}=\mathcal{O}(0.1)$. In this case, the roundtrip time of the radiation probing the boundary is shorter than the case with $\tilde{\sigma} \ll 1$, as shown in Eq.~(\ref{tauecho}).
Therefore, at early times the boundary condition affects the linear response of the object. This feature is evident in Fig.~\ref{fig:waveformlargesigma} for values of the shear viscosity $\eta\neq \eta_{\rm BH}$, for which the reflectivity of the quantum membrane is close to unity. In this case, the first GW echo interferes constructively with the prompt ringdown. 

To summarize, we have observed that for a given quantum correction $\tilde{\sigma}$, there are effects of the reflectivity of the quantum membrane that are imprinted on the GW ringdown signal.
These effects are most pronounced for $\eta \neq \eta_{\rm BH}$ but are nevertheless present in the most interesting case $\eta=\eta_{\rm BH}$, where the only difference to the classical BH case is incorporated in $\tilde{\sigma}\neq0$.

\section{
Membrane paradigm for BHs on the brane
}\label{sec:braneworld}

In Sec.~\ref{sec:quantummembrane}, we have analysed the consequences of replacing the horizon by a quantum membrane, leading to a modified boundary condition and reflectivity of the object. As a consequence, the QNM spectrum as well as the ringdown waveforms are affected. While we considered the membrane to be quantum, the background geometry remained classical, given in Eq.~(\ref{ext_metric}). In this section, we shall discuss the opposite scenario where the background spacetime inherits quantum corrections, while the membrane remains classical. Such a scenario naturally arises in the context of braneworld BHs~\cite{Emparan:1999wa,Dey:2020lhq} whose various observational signatures have been studied in~\cite{Mishra:2021waw,Banerjee:2021aln,Chakraborty:2021gdf,Dey:2020pth,Dey:2020pth,Banerjee:2019nnj,Chakraborty:2017qve,Chakravarti:2019aup}. We shall first provide the spacetime geometry of the braneworld BH, and then discuss the quantum origin of such a solution. 

\subsection{The background geometry of braneworld BHs}

The effective gravitational field equations, describing the dynamics of gravitating systems on the brane, differ from the four dimensional Einstein's field equations and receive corrections from the higher dimensional (i.e., the bulk) spacetime. In particular, the spacetime depends on the bulk Weyl tensor and quadratic combinations of the brane energy momentum tensor~\cite{Maartens:2003tw}. In the case of a vacuum brane, contributions to the effective gravitational field equations arise from the bulk Weyl tensor alone and, owing to the symmetry properties of the Weyl tensor, it behaves as the Maxwell stress-tensor with an overall \emph{negative} sign. As a consequence, the static and spherically symmetric solution takes the following form~\cite{Dadhich:2000am},
\begin{align}\label{sph_symm}
ds^{2}&=-g(r)dt^{2}+\frac{1}{g(r)}dr^{2}+r^{2}\left(d\theta^2 + \sin^2 \theta d\phi^2\right)~;
\nonumber
\\
g(r)&=1-\frac{2M}{r}+\frac{Q}{r^{2}}~,
\end{align}
where $M$ is the BH mass, $Q$ is the charge inherited from the higher dimensional Weyl tensor (which can be either positive or negative), and
$g(r)$ is a function of the radial coordinate such that one of its zeros is located at $r_{+}=M+\sqrt{M^{2}-Q}$, denoting the location of the horizon. 

Due to the presence of extra dimensions, the above BH inherits quantum corrections arising in the guise of the AdS/CFT correspondence, which conjectures that the boundary theory of a higher dimensional anti de Sitter~(AdS) spacetime is a conformal field theory (CFT). Since the bulk spacetime in the present context is AdS with some additional corrections, the BH on the brane must inherit quantum modifications due to the CFT living on the brane. Thus, the backreaction due to the CFT modifies the spacetime geometry of a braneworld BH and alters the location of the horizon to 
\begin{align}\label{Rbwbh}
R=r_{+}\left(1+\delta\right)\,,
\end{align}
where~\cite{Emparan:1999wa,Fabbri:2007kr,Dey:2020lhq}
\begin{equation}\label{delta}
    \delta \sim \frac{N^{2} l_{\rm pl}^{2}}{Mr_{+}}\,.
\end{equation}
Here $l_{\rm pl}$ is the Planck length (or the new quantum-gravity length scale), and $N$ are the CFT degrees of freedom corresponding to
\begin{align}
N^{2}\sim \left(\frac{L}{\ell_{\rm pl}} \right)^{2}\sim 10^{30}\left(\frac{L}{1\,\textrm{mm}} \right)^{2} \,,
\end{align}
where $L$ is the bulk AdS radius of the higher dimensional spacetime.
Thus, even though the membrane fluid is classical, the separation of the fluid surface from the BH horizon is due to the CFT living on the brane and hence is of purely quantum origin\footnote{In the classical limit, $\hbar\rightarrow 0$ implies $l_{\rm pl}\rightarrow 0$ leading to $\delta \rightarrow 0$ in Eq.~(\ref{delta}), as expected.}. 

The energy density and the pressure of the unperturbed membrane fluid
take the following form~\cite{Maggio:2020jml},
\begin{align} \label{rho0p0bwbh}
\rho_{0}=-\frac{\sqrt{g(R)}}{4\pi R}~;\qquad p_{0}=\frac{Rg'(R)+2g(R)}{16\pi R\sqrt{g(R)}}~,
\end{align}
%
where the derivation of the above expressions uses the Israel-Darmois junction conditions for a classical membrane~\cite{Darmois1927,Israel:1966rt,VisserBook}. 
The derivation assumes that the energy density of the membrane fluid is sufficiently small to give a negligible contribution to the bulk dynamics and hence the corrections arising from the bulk Weyl tensor can be safely ignored. In order to show that this is the case, one can compute the ratio $(\rho_{0}/\delta\textrm{Weyl})$, where $(\delta\textrm{Weyl})$ is the perturbation of the bulk Weyl tensor that scales as $\delta\textrm{Weyl} \sim (1/L)$ \cite{Kanno:2003au,Kanno:2003sc}, where $L$ is the bulk length scale. On the other hand $\rho_{0}\sim (1/M)$, where $M$ is the mass, and hence is comparable to the size of the BH. Therefore, the ratio $(\rho_{0}/\delta\textrm{Weyl})\sim (L/M)\ll 1$.
In this approximation (a very good one, since brane matter should not affect bulk dynamics) the standard GR junction conditions can be used. This is effectively equivalent to assuming continuous bulk Weyl tensor across the membrane.  

\subsection{Gravitational perturbations and boundary conditions of braneworld BHs}

\begin{figure}[t!]
\centering
\includegraphics[width=0.48\textwidth]{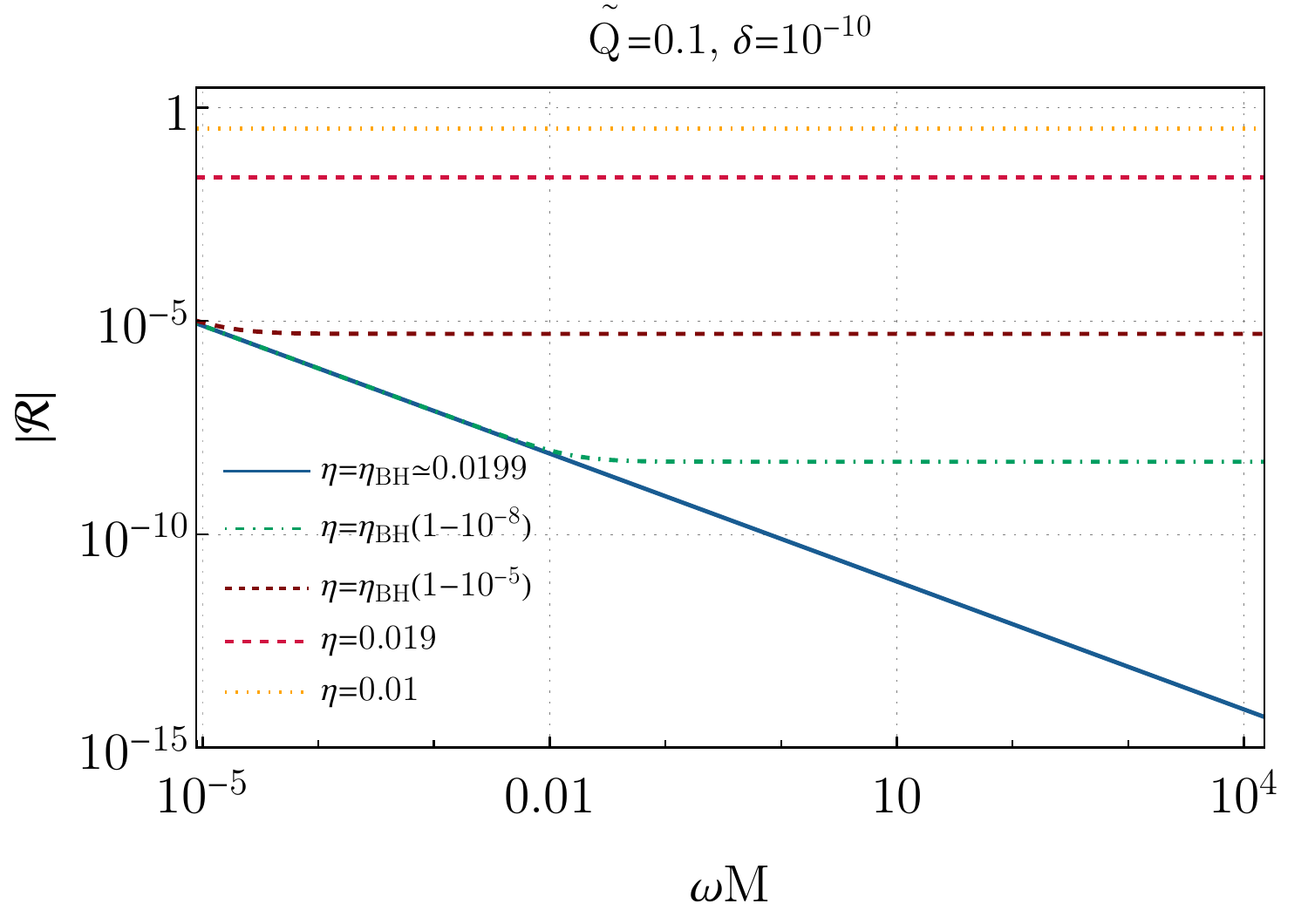}
\caption{The reflectivity of a braneworld BH as a function of the frequency for $\tilde{Q}=0.1$, $\delta=10^{-10}$, and for various choices of the membrane viscosity $\eta$. We only consider frequencies for which $\omega^2\gg V_{\rm axial}(R)$.
}
\label{fig:reflectivitysigmabwbh}
\end{figure}

Let us now discuss the linear gravitational perturbations around the background spacetime in Eq.~(\ref{sph_symm}) and derive the associated boundary condition leading to the QNMs. In particular, our aim is to understand how the presence of the quantum-origin charge $Q$ and the parameter $\delta$ affect the boundary conditions on the classical membrane and hence the ringdown. The radial part of the axial gravitational perturbation (for the polar part, see Appendix \ref{AppC}) satisfies Eq.~(\ref{ODE}) where~\cite{Toshmatov:2016bsb}
\begin{align}\label{eqnbrane}
V_{\rm axial}&=g(r)\left[\frac{\ell(\ell+1)}{r^{2}}-\frac{g'(r)}{r}-2\left(\frac{1-g(r)}{r^{2}}\right)\right]~,
\end{align}
and $x$ is the tortoise coordinate defined through $dr/dx=g(r)$. Note that the axial potential depends explicitly on the tidal charge parameter $Q$ and reduces to the standard Regge-Wheeler potential when $Q=0$. 

To solve Eq.~(\ref{ODE}) as an eigenvalue problem one requires two boundary conditions, one at infinity and another one at the location of the membrane. The boundary condition at infinity is such that the radial perturbation is a purely outgoing wave,
\begin{equation}\label{BC-infty}
\psi(r) \sim e^{i \omega x}\,, \qquad x \rightarrow \infty\,.
\end{equation}
The boundary condition on the surface $r=R$ is derived by applying the appropriate junction conditions relating the extrinsic curvature on the surface with the energy-momentum tensor of the membrane fluid. 
A straightforward extension of~\cite{Maggio:2020jml} due to the charge $Q$ and the quantum corrections through $\delta$ yields the following boundary condition,
\begin{align} \label{BCbraneworldBH}
i \omega \psi(R)&=-16\pi \eta \Bigg\{\frac{d\psi(R)}{dx}
\nonumber
\\
&+\frac{g(R)\left[\frac{\ell(\ell+1)}{R}-\frac{2(3MR-2Q)}{R^{3}} \right]}{2\left[1-\frac{3M}{R}+\frac{2Q}{R^{2}}\right]}\psi(R)\Bigg\}~,
\end{align}
which differs from the Schwarzschild horizonless case considered in \cite{Maggio:2020jml} by the nonzero values of $Q$. The case of a Schwarzschild BH, on the other hand, requires $Q$ to be set to zero as well as the quantum correction to the horizon $\delta$ to vanish. 
Note that, in the limit $R\rightarrow r_{+}$, $g(R)$ identically vanishes and the above boundary condition reduces to the boundary condition of a BH for $\eta=\eta_{\rm BH}$.

The boundary condition in Eq.~(\ref{BCbraneworldBH}) can be rewritten as

\begin{align}
\psi_{\rm M}'&=\left(-\frac{i\omega}{16\pi \eta}-\mathcal{B}\right)\psi_{\rm M}~,
\end{align}
where $\psi_{\rm M}$ is the solution of Eq.~(\ref{ODE}) near the membrane (as in Eq.~(\ref{asymptrefl})) and
the parameter $\mathcal{B}$ is defined as
\begin{align}
\mathcal{B}&\equiv \frac{g(R)\left[\frac{\ell(\ell+1)}{R}-\frac{2(3MR-2Q)}{R^{3}} \right]}{2 \left[1-\frac{3M}{R}+\frac{2Q}{R^{2}}\right]}~.
\end{align}
From Eq.~(\ref{def_reflectivity}), in the $\omega^2\gg V_{\rm axial}(R)$ regime (which holds in the relevant case when $\delta\ll1$) the reflectivity of the membrane in the background of the braneworld BH reads 
\begin{widetext}
\begin{align}
\left|\mathcal{R}\right|^{2}&=\frac{\omega^{2}\left(1-\frac{1}{16\pi \eta}\right)^{2}+\mathcal{B}^{2}}{\omega^{2}\left(1+\frac{1}{16\pi \eta}^{2}\right)+\mathcal{B}^{2}}
\nonumber
\\
&\sim \left(\frac{1-\eta/\eta_{\rm BH}}{1+\eta/\eta_{\rm BH}}\right)^{2}
+4\left(\frac{\eta}{\eta_{\rm BH}}\right)^{3}\left(\frac{1}{1+\eta/\eta_{\rm BH}}\right)^{4}
\left[\frac{2\left(\ell^2+\ell-3\right)\left(1+\sqrt{1-\tilde{Q}}\right)-\left(\ell^2+\ell-4\right)\tilde{Q}}{\left(1+\sqrt{1-\tilde{Q}}\right)^{4}}\right]^{2}\frac{\delta^{2}}{\omega^{2}M^{4}}+\mathcal{O}(\delta^{3}) \,,
\end{align}
\end{widetext}
where we defined the dimensionless quantity $\tilde{Q}=Q/M^{2}$. 
In the $\delta \rightarrow 0$ limit, $g(R)\rightarrow g(r_{+})=0$ and hence $\mathcal{B}$ identically vanishes, in which case the reflectivity depends only on the shear viscosity~\cite{Maggio:2020jml,Abedi:2020ujo}. However, for finite values of $\delta$, the reflectivity is nonzero even when $\eta=\eta_{\rm BH}$ and depends on $\tilde{Q}$, $\delta$, $\eta$ and the frequency, as shown in Fig.~\ref{fig:reflectivitysigmabwbh}.

For large frequencies, the reflectivity is given by the ratio $(\eta-\eta_{\rm BH})/(\eta+\eta_{\rm BH})$, such that for $\eta=\eta_{\rm BH}$, the reflectivity vanishes. 
This is a striking difference with respect to the quantum-membrane model of the previous section, in which the reflectivity approaches a constant value at large frequency even when $\eta=\eta_{\rm BH}$.
\subsection{QNM spectrum of the braneworld BH}

\begin{figure*}[th]
	\centering
	\includegraphics[width=0.49\textwidth]{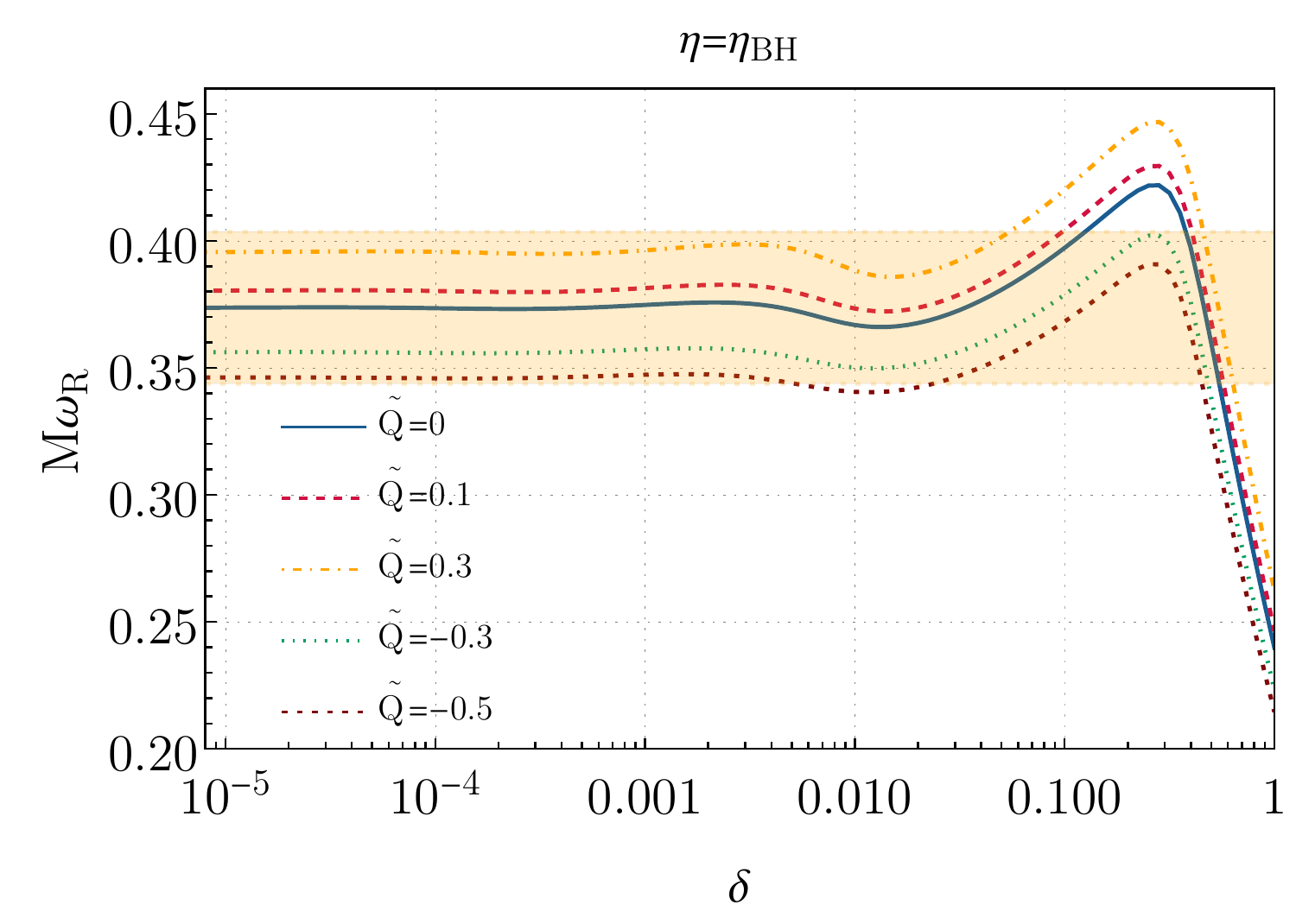}
	\includegraphics[width=0.49\textwidth]{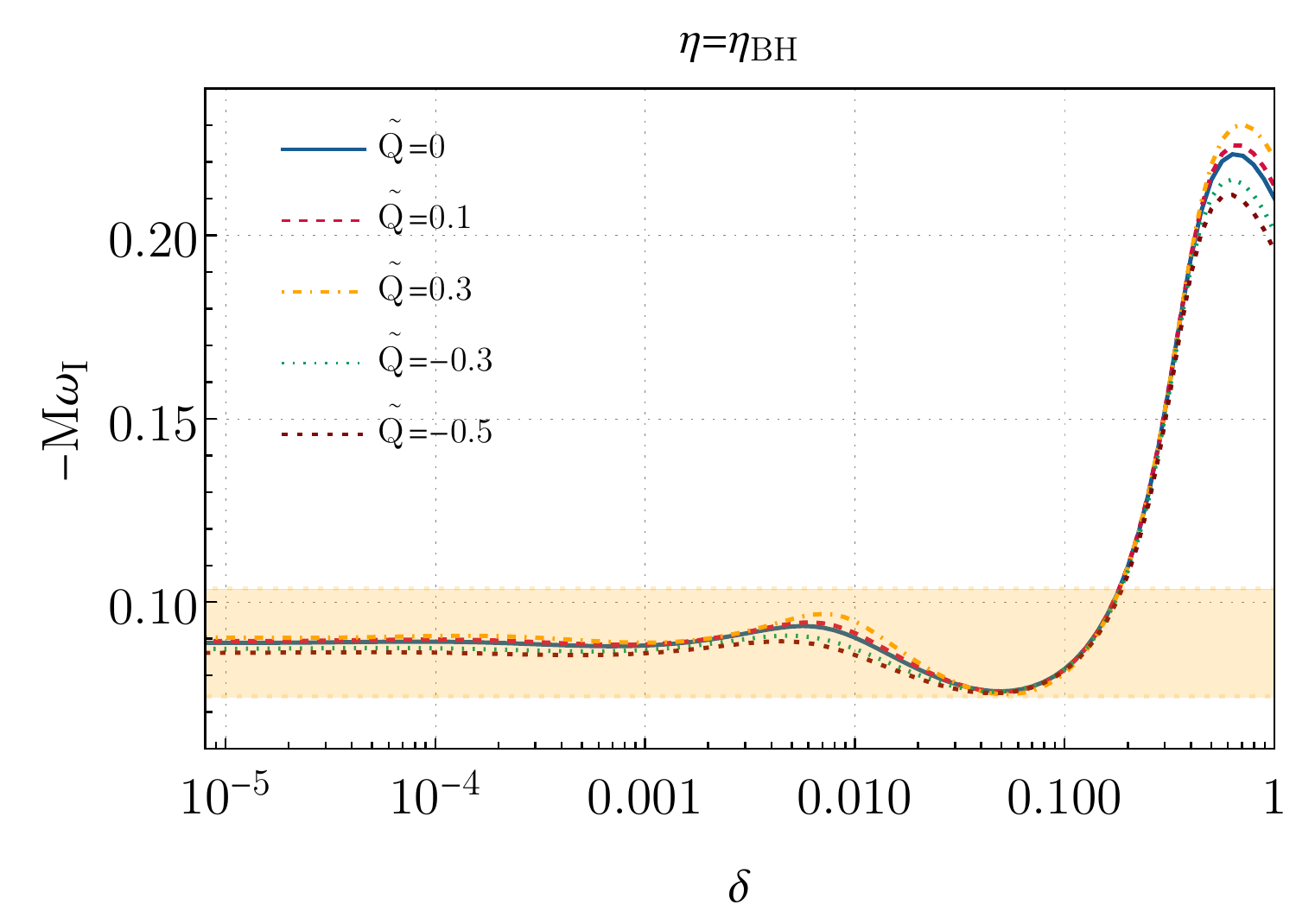}
\caption{Real (left panel) and imaginary (right panel) part of the fundamental $\ell=2$ QNM of a braneworld BH with effective shear viscosity $\eta=\eta_{\rm BH}$. The QNMs depend on the parameter $\delta$, which is related to the compactness of the object, and of the tidal charge $\tilde{Q}$. The highlighted region corresponds to the error bars associated to the fundamental $\ell=2$ QNM for the event GW150914~\cite{TheLIGOScientific:2016src,Ghosh:2021mrv}.
Horizonless objects with $\eta=\eta_{\rm BH}$, $-0.5 \lesssim \tilde{Q} \lesssim 0.3$, and $\delta \lesssim 0.05$ would be compatible with current measurements.
}
\label{fig:QNMsBWBHetabh}
\end{figure*}

Axial gravitational perturbations in the exterior of the braneworld BH are governed by Eq.~(\ref{ODE}), with the boundary conditions in Eq.~(\ref{BC-infty}) at infinity and in Eq.~(\ref{BCbraneworldBH}) on the membrane. We compute the QNMs  of the system with two numerical methods: the same continued fraction method used in the previous section and a direct integration shooting method. The former one is more robust when $\omega_{\rm I} \gtrsim \omega_{\rm R}$. When both the methods are applicable, they are in excellent agreement.

\begin{figure*}[th]
\centering
\includegraphics[width=0.49\textwidth]{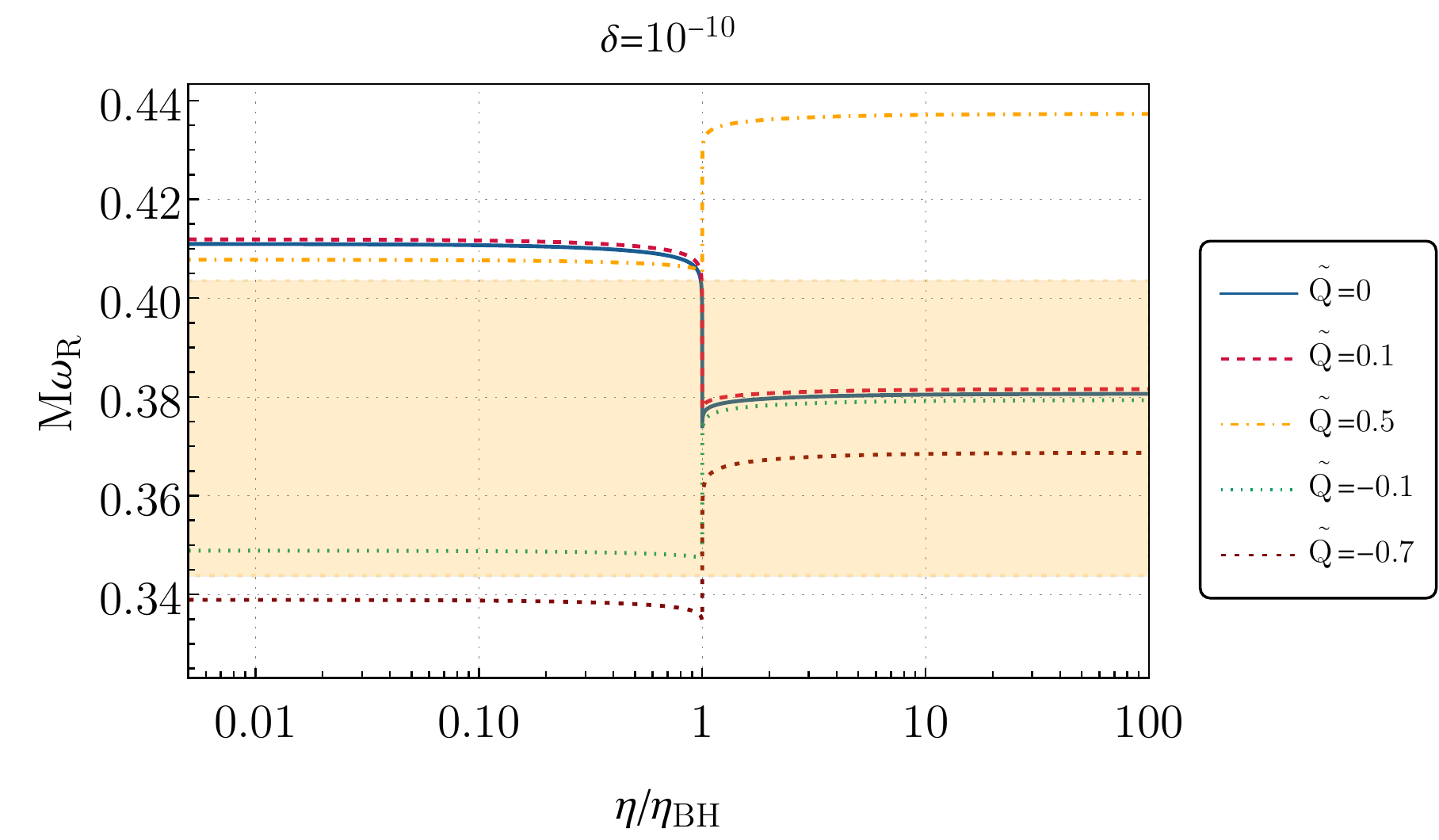}
\includegraphics[width=0.49\textwidth]{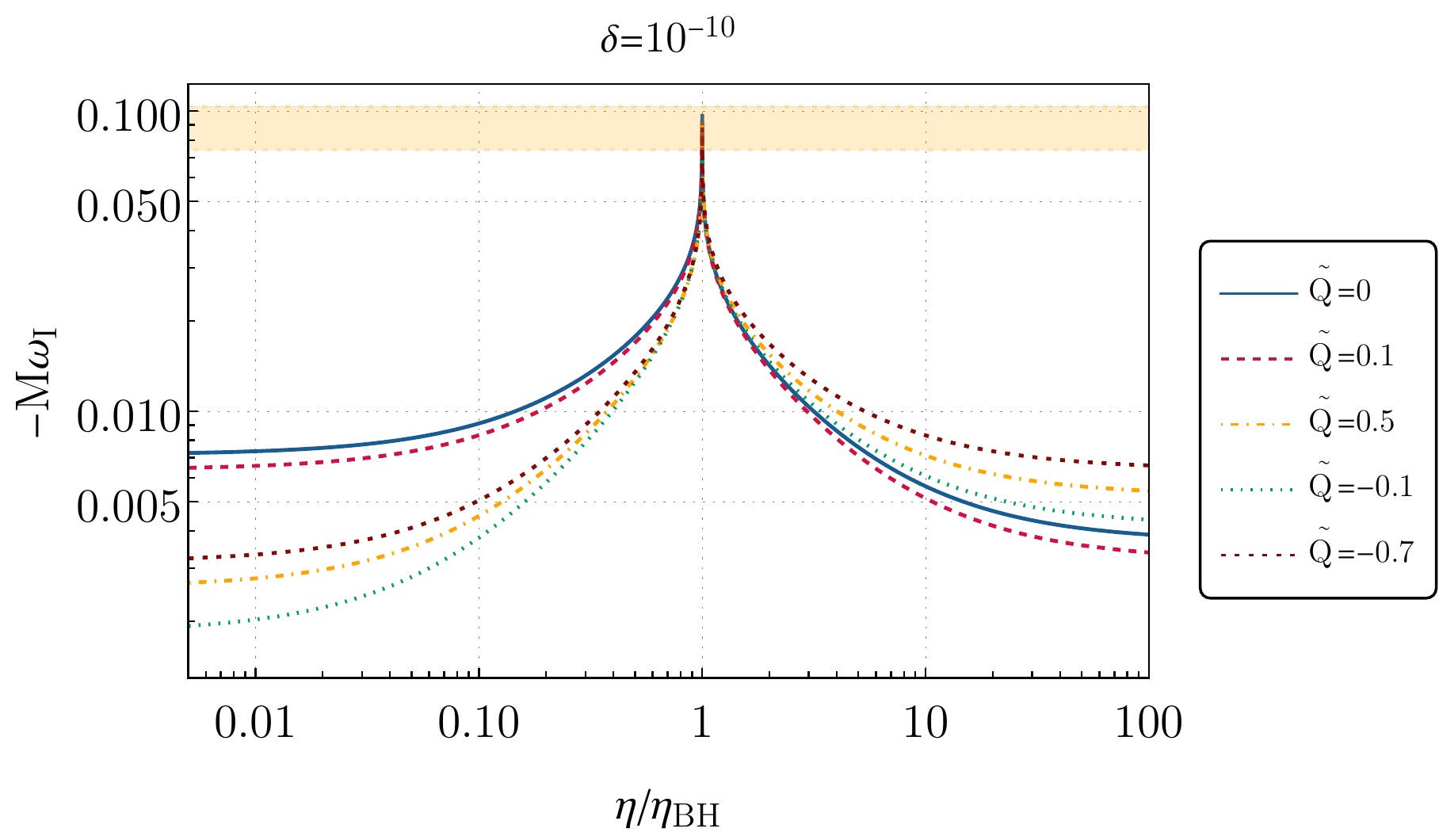}
\caption{Real (left panel) and imaginary (right panel) part of the fundamental $\ell=2$ QNM of a braneworld BH with $\delta=10^{-10}$ as a function of the effective shear viscosity $\eta$ for several values of the tidal charge $\tilde{Q}$. The highlighted region correspond to the error bars associated to the fundamental $\ell=2$ QNM for the event GW150914~\cite{TheLIGOScientific:2016src,Ghosh:2021mrv}. Values of $\eta$ slightly different from $\eta_{\rm BH}$ would be excluded by current observations regardless of the value of the tidal charge $\tilde{Q}$.
}
\label{fig:QNMsBWBH}
\end{figure*}

Let us first analyse a braneworld BH with $\eta=\eta_{\rm BH}$, for which the boundary conditions depend on $\tilde{Q}$ and $\delta$, and reduce to the Schwarzschild case as $\tilde{Q}\to0$.  The result of this analysis is presented in Fig.~\ref{fig:QNMsBWBHetabh}, where the real (left panel) and the imaginary (right panel) part of the fundamental $\ell=2$ QNM is shown as a function of $\delta$ for several values of $\tilde{Q}$, both positive and negative. As $\delta \to 0$, the QNMs tend asymptotically to a value that depend on the tidal charge $\tilde{Q}$, whereas as $\delta \gtrsim 0.01$ the QNMs start deviating from the asymptotic values. The highlighted region in Fig.~\ref{fig:QNMsBWBHetabh} corresponds to the measurement error associated to the fundamental $\ell=2$ QNM of the remnant of GW150914~\cite{TheLIGOScientific:2016src,Ghosh:2021mrv}). As Fig.~\ref{fig:QNMsBWBHetabh} explicitly shows, alternative objects with approximately $\eta=\eta_{\rm BH}$, $-0.5 \lesssim \tilde{Q} \lesssim 0.3$, and $\delta \lesssim 0.05$ would be compatible with current measurements. 

Let us now analyse the case for $\eta \neq \eta_{\rm BH}$. Fig.~\ref{fig:QNMsBWBH} shows the QNM spectrum of the braneworld BH with a quantum corrected membrane located at $\delta=10^{-10}$, as a function of the effective shear viscosity $\eta$ for several values of tidal charge $\tilde{Q}$. As $\eta$ deviates slightly from $\eta_{\rm BH}$ ($\simeq 0.02$), the QNM spectrum changes drastically. For $\eta \to 0$ and $\eta \to \infty$, the compact object is perfectly reflecting thus the boundary conditions in Eq.~(\ref{BCbraneworldBH}) reduce to Dirichlet ($\psi(R)=0$) and Neumann ($d\psi(R)/dx=0$) boundary conditions, respectively~\cite{Maggio:2020jml}. The measurement error associated to the fundamental QNM of GW150914 imposes strong constraints on the effective shear viscosity of the compact object. In particular, the behaviour of the imaginary part of the QNMs against the shear viscosity shows that, for a quantum corrected braneworld BH, values of $\eta$ slightly different from $\eta_{\rm BH}$ would be excluded regardless of the value of the tidal charge $\tilde{Q}$.

\begin{figure}[th]
\centering
\includegraphics[width=0.49\textwidth]{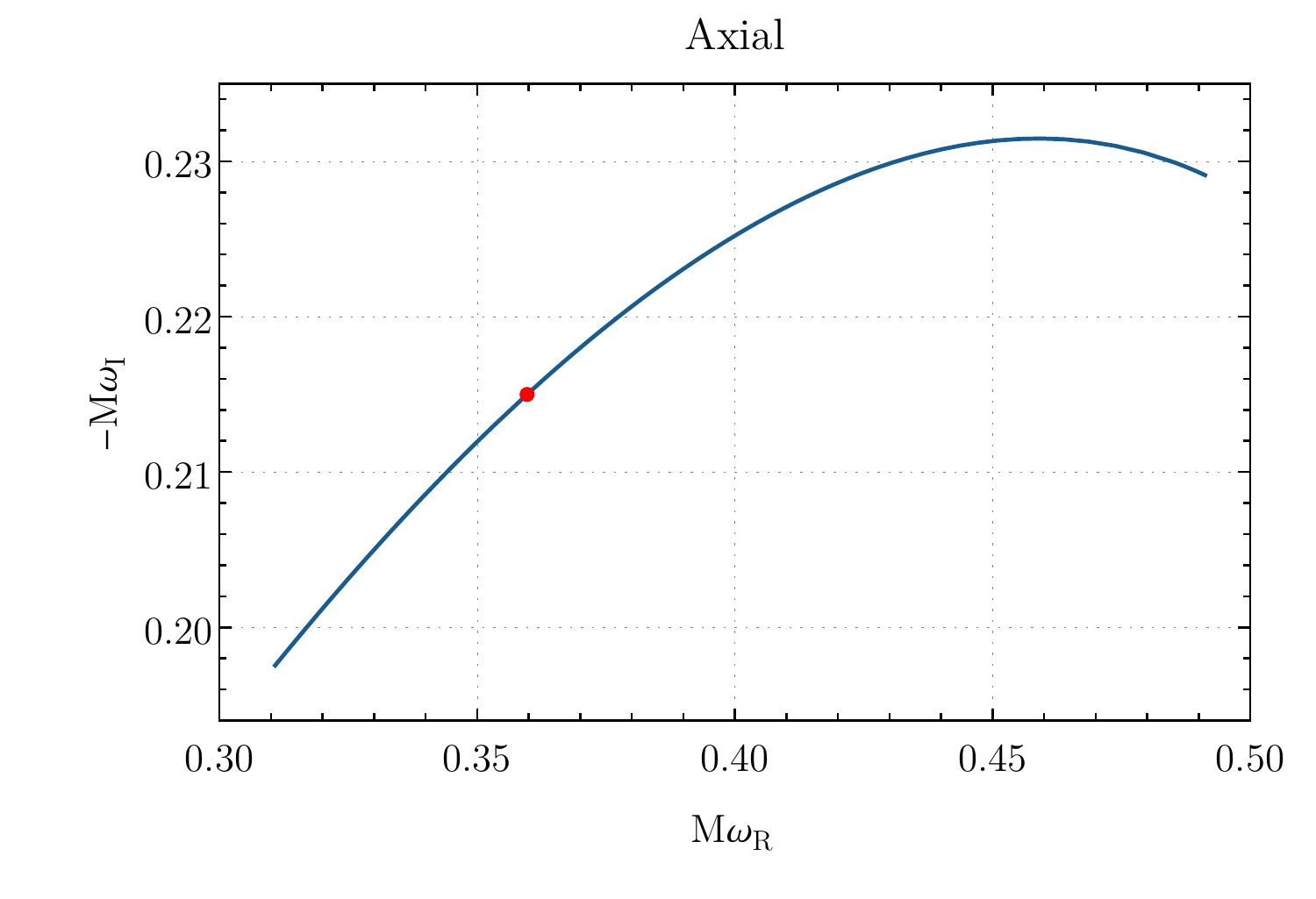}
\caption{
The complex QNM plane for the universal $\ell=2$ axial mode of a braneworld BH parametrized in terms of $\tilde{Q} \in (-1,1)$. The red marker corresponds to $Q=0$~\cite{Maggio:2020jml} and negative (positive) values of $Q$ correspond to the curve on the left (right) of the red marker.
The surface is located at the photon sphere and the mode is independent of the shear viscosity of the membrane.
}
\label{fig:universalQNM}
\end{figure}

Let us also notice that the boundary condition in Eq.~(\ref{BCbraneworldBH}) reduces to a Dirichlet boundary condition when the radius of the compact object is located at the photon sphere, $r_{\rm ph}=(1/2)[3M+\sqrt{9M^2-8Q}]$. In this case, the axial QNM spectrum is universal regardless of the effective shear viscosity of the quantum membrane, as shown in Fig.~\ref{fig:universalQNM}.
In this plot, we show the real and the imaginary part of the universal mode as parametric functions of $\tilde{Q} \in (-1,1)$. 
The detection of this peculiar mode would be a clear signature of the location of the radius of the object regardless of its reflectivity.

At this outset, let us briefly discuss the difference of the approach presented here to the one considered in \cite{Rahman:2021kwb}. Even though the background metric looks similar, their origins are different. In our case, the term $Q$ origins from braneworld and can be either positive or negative, while in \cite{Rahman:2021kwb} the term $Q^{2}$ origins from electric charge and is strictly positive. As a consequence, the equations governing the gravitational perturbations and hence the boundary conditions are also very different between the one presented here and in \cite{Rahman:2021kwb}. Moreover, the distance between the horizon and the membrane in our case is purely quantum, arising due to CFT, while in \cite{Rahman:2021kwb} is an assumption of the model.

\section{Discussion and Concluding Remarks}\label{sec:conclusions}

Exploring the BH nature of ultracompact objects and their theoretical and observational consequences is one of the key ingredients of present day GW research. In searching for ultracompact objects other than BHs and neutron stars, one often considers the existence of some exotic matter or invokes cutoff surfaces near the would-be horizon in an ad-hoc manner. In this work, we have provided a natural origin for horizonless ultracompact objects by invoking quantum effects. Considering the BH horizon being replaced by a fluid consisting of some quantum harmonic oscillators in the ground state, we showed that the horizon gets replaced by a quantum membrane with \emph{non-zero} and \emph{frequency-dependent} reflectivity. The junction condition on the membrane also gets modified by the existence of the quantum fluid, leading to a distinct boundary condition for gravitational perturbations on the membrane. The boundary condition depends on the quantum nature of the membrane in an explicit manner. As a consequence, in this setup the reflectivity of the membrane becomes nonzero only in the presence of quantum effects, while it identically vanishes in the classical BH limit. This leads to significant changes in the QNM spectrum, as well as in the time-domain ringdown waveform. 
The latter contains echoes originating from the nonvanishing membrane reflectivity, even when the shear viscosity equals that of a classical BH, $\eta=\eta_{\rm BH}$. However, the effective reflectivity is small so that the echo amplitude is smaller than in some toy models in which the reflectivity is close to unity.
In our basic model the shear viscosity is a free parameter, which we often set to the classical BH value to minimize the number of new free parameters in the model. An interesting future extension could be to find an explicit expression for the shear viscosity in terms of the quantum properties of the membrane.
Another interesting example is to explicitly compute the stimulated Hawking emission from the quantum membrane, along the lines of~\cite{Oshita:2019sat,Wang:2019rcf,Abedi:2021tti}.

As another example of the quantum membrane, we considered the case of braneworld BH where through the AdS/CFT correspondence one may argue that a CFT should be present on the brane, acting as a natural candidate for the quantum fluid. Thus, the braneworld BH receives a contribution from the extra dimensions, manifesting as a tidal charge parameter $\tilde{Q}$, as well as from CFT, leading to a reflective membrane away from the horizon. The departure from the horizon is again due to quantum effects of the CFT, leading to the existence of a quantum membrane similar to the previous case. As a consequence, the boundary condition for the gravitational perturbation near the horizon are modified, affecting the QNM spectrum and the ringdown signal. 
In particular, in order for the real part of the QNMs to be consistent with the upper bound imposed by GW150914, the tidal charge parameter $Q$ must lie within the range $-0.5<\tilde{Q}<0.3$, as well as the departure from the horizon should satisfy $\delta<0.05$. This leads to the following bound on the CFT degrees of freedom on the brane, $N^{2}<10^{-1}(Mr_{+}/l_{\rm pl}^{2})$. 
Overall, we observe that the quantum effects can naturally lead to the QNM spectrum and the time-domain waveform to be different from that of BHs, with the presence of echoes due to the finite reflectivity of the quantum membrane. Since the above analysis is for a static and spherically symmetric spacetime, the constraints arising from GW observations are not directly applicable. The generalization of the above setup to spinning objects is an important extension that we leave for future work.        

\section*{Acknowledgements}

The authors are grateful to Luca Buoninfante for useful discussions.
Research of S.C. is funded by the INSPIRE Faculty fellowship from the DST, Government of India (Reg. No. DST/INSPIRE/04/2018/000893) and by the Start-Up Research Grant from SERB, DST, Government of India (Reg. No. SRG/2020/000409). P.P. and E.M. acknowledge financial support provided under the European Union's H2020 ERC, Starting Grant agreement no.~DarkGRA--757480, under the MIUR PRIN and FARE programmes (GW-NEXT, CUP:~B84I20000100001), and support from the Amaldi Research Center funded by the MIUR program ``Dipartimento di Eccellenza" (CUP:~B81I18001170001). 
AM’s research is funded by the Netherlands Organisation for Science and Research (NWO) grant number 680-91-119.
\appendix
\begin{widetext}
\section{Properties of the quantum membrane away from the horizon}\label{AppA}

In this appendix, we discuss the properties of a quantum membrane consisting of some harmonic oscillators away from the horizon by a distance $\upsilon$. The case discussed in the main text, where the membrane sits on the would-be horizon, is recovered when $\upsilon\to0$. 

The normalization integral for the ground state wave function yields
\begin{align}
\int_{-\upsilon}^{\infty} d\epsilon |\Psi(\epsilon)|^{2}&=|A|^{2}\int_{-\upsilon}^{\infty} d\epsilon~\exp(-\epsilon^{2}/\sigma^{2})
\nonumber
\\
&=|A|^{2}\left(\frac{\sigma}{2}\right)\int_{\upsilon^{2}/\sigma^{2}}^{\infty}dz~z^{-1/2}e^{-z}
=|A|^{2}\left(\frac{\sigma}{2}\right)\Gamma\left(\frac{1}{2},\frac{\upsilon^{2}}{\sigma^{2}}\right)~,
\end{align}
where $\Gamma(a,x)$ is the incomplete Gamma function. Thereby we fix the normalization constant,
\begin{align}\label{CaseI_norm}
|A|^{2}=\left(\frac{2}{\sigma}\right)\frac{1}{\Gamma\left(\frac{1}{2},\frac{\upsilon^{2}}{\sigma^{2}}\right)}~.
\end{align}
Furthermore, the expectation value of $\hat{\epsilon}$ reads
\begin{align}
\langle \hat{\epsilon}\rangle&=\int_{-\upsilon}^{\infty} d\epsilon~ \epsilon |\Psi(\epsilon)|^{2}
=|A|^{2}\left(\frac{\sigma^{2}}{2}\right)\int_{\upsilon^{2}/\sigma^{2}}^{\infty}dz~e^{-z}
=\frac{\sigma}{\Gamma\left(\frac{1}{2},\frac{\upsilon^{2}}{\sigma^{2}}\right)}e^{-\frac{\upsilon^{2}}{\sigma^{2}}}~.
\end{align}
As expected, one recovers Eq.~(\ref{exp_epsilon}) by taking the $\upsilon\to0$ limit of the above expression. Thus, the classical location of the fluid surface in the scenario considered here becomes, 
\begin{align}\label{position_membrane}
R=r_{+}+\upsilon+\langle \hat{\epsilon}\rangle =r_{+}+\upsilon+\frac{\sigma}{\Gamma\left(\frac{1}{2},\frac{\upsilon^{2}}{\sigma^{2}}\right)}e^{-\frac{\upsilon^{2}}{\sigma^{2}}}~,
\end{align}
Finally, the quantum nature of the membrane is captured by the following expression,  
\begin{align}\label{qdiff_caseI}
\langle \hat{\epsilon}^{2}\rangle-\langle \hat{\epsilon}\rangle^{2}=\frac{\sigma^{2}}{\Gamma\left(\frac{1}{2},\frac{\upsilon^{2}}{\sigma^{2}}\right)}\left[\Gamma\left(\frac{3}{2},\frac{\upsilon^{2}}{\sigma^{2}}\right)-\frac{e^{-2\frac{\upsilon^{2}}{\sigma^{2}}}}{\Gamma\left(\frac{1}{2},\frac{\upsilon^{2}}{\sigma^{2}}\right)}\right]~.
\end{align}
The expectation value of the relevant components of the stress-energy tensor yields
\begin{align}
\langle \hat{T}_{tt}\rangle&=\rho_{0}\langle f(r_{0}+\hat{\epsilon})\rangle
=\rho_{0} \left[f(r_{0})+f'(r_{0})\langle \hat{\epsilon}\rangle + \frac{1}{2}f''(r_{0})\langle \hat{\epsilon}^{2}\rangle +\mathcal{O}(\tilde{\sigma}^{3}) \right]
\nonumber
\\
&=\rho_{0}\left[f(R)+\frac{1}{2}f''(r_{0})\left\{\langle \hat{\epsilon}^{2}\rangle-\langle \hat{\epsilon}\rangle^{2} \right\}+\mathcal{O}(\tilde{\sigma}^{3})\right]~,
\end{align}
as well as
\begin{align}
\langle \hat{T}_{\theta \theta}\rangle&=p_{0}\langle \left(r_{0}+\hat{\epsilon}\right)^{2}\rangle
=p_{0}\left(r_{0}^{2}+2r_{0}\langle \hat{\epsilon}\rangle+\langle \hat{\epsilon}^{2}\rangle \right)
\nonumber
\\
&=p_{0}\left(R\right)^{2}+p_{0}\left\{\langle \hat{\epsilon}^{2}\rangle-\langle \hat{\epsilon}\rangle^{2} \right\}
=\frac{\langle \hat{T}_{\phi \phi}\rangle}{\sin^{2}\theta}~,
\end{align}
where $r_{0}=r_{+}+\upsilon$. Finally, from the semiclassical junction condition given in Eq.~(\ref{semi_class_junc}), we obtain the energy density of the quantum fluid living on the membrane,
\begin{align}
\rho_{0}&=-\frac{f(R)^{3/2}}{4\pi R}\frac{1}{f(R)+\frac{1}{2}f''(r_{0})\left(\langle \hat{\epsilon}^{2}\rangle-\langle \hat{\epsilon}\rangle^{2} \right)+\mathcal{O}(\tilde{\sigma}^{3})}~,
\end{align}
while the expression for the pressure is
\begin{align}
p_{0}&=\frac{R\left[2f(R)+Rf'(R)\right]}{16\pi\sqrt{f(R)}}\frac{1}{R^{2}+\left(\langle \hat{\epsilon}^{2}\rangle-\langle \hat{\epsilon}\rangle^{2} \right)}~.
\end{align}
Note that, in the limit $\upsilon \rightarrow 0$, the expressions for $\rho_{0}$ and $p_{0}$ coincides with the expressions in Sec.~\ref{sec:quantummembrane}. Given the above expressions for the energy density and pressure, we obtain the following expression for the combination, 

\begin{align}
\rho_{0}+p_{0}&=\frac{1}{16\pi R\sqrt{f(R)}\left[f(R)+\frac{1}{2}f''(r_{0})\left(\langle \hat{\epsilon}^{2}\rangle-\langle \hat{\epsilon}\rangle^{2} \right)+\mathcal{O}\left(\tilde{\sigma}^{3}\right)\right]\left[R^{2}+\left(\langle \hat{\epsilon}^{2}\rangle-\langle \hat{\epsilon}\rangle^{2} \right) \right]}
\nonumber
\\
&\times \Bigg\{-4f(R)^{2}\left[R^{2}+\left(\langle \hat{\epsilon}^{2}\rangle-\langle \hat{\epsilon}\rangle^{2} \right) \right]
\nonumber
\\
&\hskip 1 cm +R^{2}\left[2f(R)+Rf'(R) \right]\left[f(R)+\frac{1}{2}f''(r_{0})\left(\langle \hat{\epsilon}^{2}\rangle-\langle \hat{\epsilon}\rangle^{2} \right)+\mathcal{O}\left(\tilde{\sigma}^{3}\right)\right]\Bigg\}
\nonumber
\\
&=\frac{R^{2}f(R)\left[-2f(R)+Rf'(R)\right]}
{16\pi R\sqrt{f(R)}\left\{f(R)R^{2}+\left[\frac{1}{2}f''(r_{0})R^{2}+f(R) \right]\left(\langle \hat{\epsilon}^{2}\rangle-\langle \hat{\epsilon}\rangle^{2} \right)+\mathcal{O}\left(\tilde{\sigma}^{3}\right)\right\}}
\nonumber
\\
&+\frac{\left(\langle \hat{\epsilon}^{2}\rangle-\langle \hat{\epsilon}\rangle^{2} \right)\left\{-4f(R)^{2}+\frac{1}{2}f''(r_{0})R^{2}\left[2f(R)+Rf'(R) \right] \right\}}
{16\pi R\sqrt{f(R)}\left\{f(R)R^{2}+\left[\frac{1}{2}f''(r_{0})\left(R\right)^{2}+f(R) \right]\left(\langle \hat{\epsilon}^{2}\rangle-\langle \hat{\epsilon}\rangle^{2} \right)+\mathcal{O}\left(\tilde{\sigma}^{3}\right)\right\}} \,.
\end{align}
Note that in the absence of any quantum corrections, or of a classical stochastic treatment of the membrane, we have $\langle \hat{\epsilon}^{2}\rangle-\langle \hat{\epsilon}\rangle^{2}=0$, along with $R=r_{0}$, yielding, 
\begin{align}
\rho_{0}+p_{0}=\frac{-2f(r_{0})+r_{0}f'(r_{0})}{16\pi r_{0}\sqrt{f(r_{0})}}~,
\end{align}
which coincides identically with the result derived in~\cite{Maggio:2020jml}. 
 
We now present the discussion for axial perturbations of the gravitational sector. First of all, the perturbations of the extrinsic curvature components remain identical to those in Sec.~\ref{sec:quantummembrane}, whereas the components of the stress energy tensor becomes
\begin{align}
\langle\delta \hat{T}_{t\phi}\rangle&=-e^{-i\omega t}\rho_{0}(R)h_{0}(R)\sin \theta~\partial_{\theta}P_{\ell}(\cos\theta)-\sin^{2}\theta \delta u^{\phi}\left(\rho_{0}+p_{0}\right)\langle~ \hat{R}^{2}\sqrt{f(\hat{R})}~\rangle \,,
\\
\langle\delta \hat{T}_{\theta \phi}\rangle&=-\eta \sin^{2}\theta \partial_{\theta}\delta u^{\phi}\langle \hat{R}^{2}\rangle
=-\eta \sin^{2}\theta \partial_{\theta}\delta u^{\phi}\left[R^{2}+\left(\langle \hat{\epsilon}^{2}\rangle-\langle \hat{\epsilon}\rangle^{2} \right)\right]\,.
\end{align}
From the perturbed semiclassical junction condition $\delta K_{ab}-K\delta h_{ab}=-8\pi\langle \delta \hat{T}_{ab}\rangle$, we obtain from the $(t,\phi)$ component,
\begin{align}\label{deltau}
\delta u^{\phi}&=\frac{e^{-i\omega t}\partial_{\theta}P_{\ell}(\cos\theta)}{8\pi \sin\theta \left(\rho_{0}+p_{0}\right)\langle~ \hat{R}^{2}\sqrt{f(\hat{R})}~\rangle}
\Bigg[\frac{1}{2}\sqrt{f(R)}\left[i\omega h_{1}(R)+h_{0}'(R)\right]
\nonumber
\\
&\hskip 1 cm -\left(\frac{f'(R)}{2\sqrt{f(R)}}+\frac{2\sqrt{f(R)}}{R}\right)h_{0}(R)-8\pi \rho_{0}(R)h_{0}(R)\Bigg]~.
\end{align}
Along identical lines, the semiclassical junction condition associated with the $(\theta,\phi)$ component is,
\begin{align}
-\frac{1}{2}e^{-i\omega t}\sqrt{f(R)}&h_{1}(R)\left[-\cos\theta~\partial_{\theta}P_{\ell}(\cos\theta)+ \sin \theta~\partial_{\theta}^{2}P_{\ell}(\cos\theta)\right]
\nonumber
\\
&=8\pi \eta \sin^{2}\theta \partial_{\theta}\delta u^{\phi}\left[R^{2}+\left(\langle \hat{\epsilon}^{2}\rangle-\langle \hat{\epsilon}\rangle^{2} \right)\right]~.
\end{align}
Substitution of $\delta u^{\phi}$ from Eq.~(\ref{deltau}) yields the following expression for the perturbation $h_{1}(R)$, 
\begin{align}
-\frac{1}{2}\sqrt{f(R)}~h_{1}(R)&=\frac{\eta\left[R^{2}+\left(\langle \hat{\epsilon}^{2}\rangle-\langle \hat{\epsilon}\rangle^{2} \right)\right]}{\left(\rho_{0}+p_{0}\right)\langle~ \hat{R}^{2}\sqrt{f(\hat{R})}~\rangle}
\Bigg[\frac{1}{2}\sqrt{f(R)}\left[i\omega h_{1}(R)+h_{0}'(R)\right]
\nonumber
\\
&\hskip 1 cm -\left(\frac{f'(R)}{2\sqrt{f(R)}}+\frac{2\sqrt{f(R)}}{R}\right)h_{0}(R)-8\pi \rho_{0}(R)h_{0}(R)\Bigg]~.
\end{align}
Introducing the Regge-Wheeler function, we obtain the final form of the boundary condition on the stretched horizon, located classically at $r=R=r_{+}+\upsilon+\langle \hat{\epsilon}\rangle$:
\begin{align}
i\omega~\psi(R)&=\frac{\eta\left[R^{2}+\left(\langle \hat{\epsilon}^{2}\rangle-\langle \hat{\epsilon}\rangle^{2} \right)\right]}{\left(\rho_{0}+p_{0}\right)\langle~ \hat{R}^{2}\sqrt{f(\hat{R})}~\rangle}
\Bigg[V_{\rm axial}(R)\psi(R)-\frac{1}{R}\frac{d\psi(R)}{dx}\left[Rf'(R)-2f(R)\right]
\nonumber
\\
&-\frac{4f(R)}{R}\left(\frac{d\psi(R)}{dx}+\frac{f(R)}{R}\psi(R)\right)\left(1+\frac{4\pi \rho_{0}(R)R}{\sqrt{f(R)}}\right)\Bigg]~.
\end{align}
In the limit of vanishing quantum corrections, by using the expression for $\rho_0(R)$, it can be shown that the term within the round bracket in the last line identically vanishes, thus recovering the result derived in~\cite{Maggio:2020jml}.   

\section{Polar perturbation for the quantum membrane}\label{AppB}

In the main text we have discussed in detail the axial gravitational perturbations, since the polar case is much more involved, which would have taken us away from the central ideas of the model. However, for completeness, in what follows we shall discuss the basic ingredients for the derivation of the polar gravitational perturbations for the quantum membrane. 

For polar gravitational perturbations, the metric perturbations are given by the following expressions~\cite{Maggio:2020jml},
\begin{align}
\delta g_{tt}&=e^{-i\omega t}P_{\ell}(\cos \theta)f(r)\mathcal{H}(r)~;
\label{met_pol_tt}
\\
\delta g_{rr}&=e^{-i\omega t}P_{\ell}(\cos \theta)\frac{\mathcal{H}(r)}{f(r)}~;
\label{met_pol_rr}
\\
\delta g_{tr}&=e^{-i\omega t}P_{\ell}(\cos \theta)\mathcal{H}_{1}(r)~;
\label{met_pol_tr}
\\
\delta g_{\theta \theta}&=e^{-i\omega t}P_{\ell}(\cos \theta)r^{2}\mathcal{K}(r)~.
\label{met_pol_theta}
\end{align}
Thus, the metric perturbations in the polar sector are given by three unknown functions, $\mathcal{H}(r)$, $\mathcal{H}_{1}(r)$, and $\mathcal{K}(r)$. In the polar case, the perturbed components of the normal vector are
\begin{align}
\delta n_{t}&=\frac{1}{\sqrt{f}}\epsilon \left(i\omega\right)e^{-i\omega t}P_{\ell}(\cos \theta)\delta R_{0}~;
\label{norm_polar_t}
\\
\delta n_{r}&=\frac{1}{2\sqrt{f}} e^{-i\omega t}P_{\ell}(\cos \theta)\mathcal{H}(r)~;
\label{norm_polar_r}
\\
\delta n_{\theta}&=-\frac{1}{\sqrt{f}}\epsilon e^{-i\omega t}\delta R_{0}\partial_{\theta}P_{\ell}(\cos \theta)~,
\label{norm_polar_theta}
\end{align}
where $\delta R_{0}$ corresponds to the shift in the classical location of the membrane due to the polar gravitational perturbations. Along with these, the perturbations of the extrinsic curvature components follow from~\cite{Maggio:2020jml} yielding,
\begin{align}
\delta K_{t\theta}&=-\frac{1}{2\sqrt{f}} e^{-i\omega t}\partial_{\theta}P_{\ell}(\cos \theta)\left(f\mathcal{H}_{1}-2i\omega \delta R_{0} \right)~,
\label{pert_K_ttheta}
\\
\delta K_{tt}&=\frac{1}{4\sqrt{f}} e^{-i\omega t}P_{\ell}(\cos \theta)\Big[\delta R_{0}\left(4\omega^{2}-f'^{2}\right)+2f^{2}\mathcal{H}'+f\left(4i\omega \mathcal{H}_{1}-2f''\delta R_{0}+3f'\mathcal{H}\right)\Big]~,
\label{pert_K_tt}
\\
\delta K_{\theta \theta}&=\frac{e^{-i\omega t}}{2\sqrt{f}}\Big[f\left(2\delta R_{0}-R\mathcal{H}+R^{2}\mathcal{K}'+2R\mathcal{K}\right)+\delta R_{0}\left(Rf'-2\partial_{\theta}^{2}\right)\Big]P_{\ell}(\cos \theta)~,
\label{pert_K_theta}
\\
\delta K_{\phi \phi}&=\frac{e^{-i\omega t}}{2\sqrt{f}}\sin^{2}\theta\Big[f\left(2\delta R_{0}-R\mathcal{H}+R^{2}\mathcal{K}'+2R\mathcal{K}\right)+\delta R_{0}\left(Rf'-\cot \theta \partial_{\theta}\right)\Big]P_{\ell}(\cos \theta)~.
\label{pert_K_phi}
\end{align}
All of these expressions are evaluated on the surface $r=R$. Until this point, the computation is purely on the geometry side and hence classical. The effect of the quantum fluid of the membrane arises from the perturbed stress-energy tensor components, whose expectations values is on the right hand side of the semiclassical junction condition. The expectation values of the perturbed energy momentum tensors yield
\begin{align}
\langle \delta T_{tt}\rangle &=e^{-i\omega t}P_{\ell}(\cos \theta)\left[\rho_{0} \delta R_{0}\langle f'(r_{+}+\hat{\epsilon}) \rangle
+\left(\rho_{1}-\rho_{0}\mathcal{H}\right)\langle f(r_{+}+\hat{\epsilon}) \rangle\right]
\nonumber
\\
&=e^{-i\omega t}P_{\ell}(\cos \theta)\Bigg[\rho_{0} \delta R_{0}\left\{f'(R)+\frac{1}{2}f'''(r_{+})\left(\langle \hat{\epsilon}^{2} \rangle -\langle \hat{\epsilon} \rangle^{2}\right)\right\}
\nonumber
\\
&\qquad +\left(\rho_{1}-\rho_{0}\mathcal{H}\right)\left\{f(R)+\frac{1}{2}f''(r_{+})\left(\langle \hat{\epsilon}^{2} \rangle -\langle \hat{\epsilon} \rangle^{2}\right)\right\}\Bigg]~,
\\
\langle \delta T_{\theta \theta}\rangle &=-\langle R^{2} \rangle \left[\left(\zeta+\eta\right)\partial_{\theta}\delta u^{\theta}+\left(\zeta-\eta\right) \cot \theta \delta u^{\theta}\right]
\nonumber
\\
&\qquad +e^{-i\omega t}P_{\ell}(\cos \theta)\left[\left(p_{1}+\mathcal{K}p_{0}\right)\langle R^{2} \rangle+2p_{0}\delta R_{0} \langle R \rangle+i\zeta \omega \mathcal{K}\langle \frac{R^{2}}{\sqrt{f}}\rangle +2i\zeta \omega \delta R_{0} \langle \frac{R}{\sqrt{f}}\rangle\right]
\nonumber
\\
&=
-\langle R^{2} \rangle \left[\left(\zeta+\eta\right)\partial_{\theta}\delta u^{\theta}+\left(\zeta-\eta\right) \cot \theta \delta u^{\theta}\right]
\nonumber
\\
&\qquad +e^{-i\omega t}P_{\ell}(\cos \theta)\left[\left(p_{1}+\mathcal{K}p_{0}\right)\langle R^{2} \rangle+2p_{0}\delta R_{0} \langle R \rangle+i\zeta \omega \frac{\mathcal{K}}{\sqrt{f(R)}}\langle R^{2}\rangle +2i\zeta \omega \frac{\delta R_{0}}{\sqrt{f(R)}} \langle R\rangle\right]~,
\\
\langle \delta T_{\phi \phi}\rangle &=-\langle R^{2} \rangle \sin^{2}\theta \left[\left(\zeta-\eta\right)\partial_{\theta}\delta u^{\theta}+\left(\zeta+\eta\right) \cot \theta \delta u^{\theta}\right]
\nonumber
\\
&\qquad +e^{-i\omega t}P_{\ell}(\cos \theta)\left[\left(p_{1}+\mathcal{K}p_{0}\right)\langle R^{2} \rangle+2p_{0}\delta R_{0} \langle R \rangle+i\zeta \omega \mathcal{K}\langle \frac{R^{2}}{\sqrt{f}}\rangle +2i\zeta \omega \delta R_{0} \langle \frac{R}{\sqrt{f}}\rangle\right]
\nonumber
\\
&=-\langle R^{2} \rangle \sin^{2}\theta \left[\left(\zeta-\eta\right)\partial_{\theta}\delta u^{\theta}+\left(\zeta+\eta\right) \cot \theta \delta u^{\theta}\right]
\nonumber
\\
&\qquad +e^{-i\omega t}P_{\ell}(\cos \theta)\left[\left(p_{1}+\mathcal{K}p_{0}\right)\langle R^{2} \rangle+2p_{0}\delta R_{0} \langle R \rangle+i\zeta \omega \frac{\mathcal{K}}{\sqrt{f(R)}}\langle R^{2}\rangle +2i\zeta \omega \frac{\delta R_{0}}{\sqrt{f(R)}} \langle R\rangle\right]~,
\\
\langle \delta T_{t \theta}\rangle &=-\langle R^{2} \sqrt{f}\rangle \left(\rho_{0}+p_{0}\right)\delta u^{\theta}
\nonumber
\\
&=-\langle R^{2} \rangle \sqrt{f(R)} \left(\rho_{0}+p_{0}\right)\delta u^{\theta}~,
\end{align}
where the first-order corrections of the energy density and the pressure of the membrane are $\delta \rho=e^{-i\omega t}P_{\ell}(\cos \theta)\rho_{1}(r)$ and $\delta p=e^{-i\omega t}P_{\ell}(\cos \theta)p_{1}(r)$. 
Note that when the expectation values of nonpositive integer powers of $f(\hat{R})$ are present, we ignore the operator nature and approximate them to be functions of $f(R)$. The expectation values of $\hat{R}$ and $\hat{R}^{2}$ appearing in the above expressions are easy to obtain using the expression $\hat{R}=R+\hat{\epsilon}$ and then computing the expectation values of $\hat{\epsilon}$ and $\hat{\epsilon}^{2}.$

Using the perturbed version of the semiclassical junction condition, and the expressions for $\delta K_{ab}$ and $\langle \delta \hat{T}_{ab}\rangle$, one may obtain $\delta R_{0}$, $\rho_{1}$, and $p_{1}$. These relations, along with the Zerilli potential for the spherically symmetric background spacetime, will provide the desired boundary conditions for polar perturbations, which will depend on the quantum properties of the membrane. 

\section{Polar perturbation for braneworld BHs: Zerilli potential}\label{AppC}

In the main text, we have discussed the axial perturbations of the braneworld BH. In this appendix, we also present the polar perturbations for completeness. The analysis of polar perturbation will follow the same lines as in~\cite{Maggio:2020jml}, with nonzero metric perturbations being given by Eq.~(\ref{met_pol_tt}) to Eq.~(\ref{met_pol_theta}). For the braneworld BH as well, the polar perturbation will shift the location of the membrane at $R=r_{+}+\delta$ to $R+\delta R_{0}$, such that the expression for the components of the perturbed normal vector becomes those given in Eq.~(\ref{norm_polar_t}) to Eq.~(\ref{norm_polar_theta}). Similarly, the perturbed extrinsic curvature components are those given by Eq.~(\ref{pert_K_ttheta}) to Eq.~(\ref{pert_K_phi}). The perturbed energy momentum tensor takes the form as in~\cite{Maggio:2020jml}. 

The only difference being the dependence of the Zerilli potential~\cite{Zerilli:1970se,Zerilli:1971wd} on the tidal charge $Q$, due to the presence of the extra dimensions. First of all, one notices that the three perturbation variables $\mathcal{H}(r)$, $\mathcal{H}_{1}(r)$, and $\mathcal{K}(r)$ are related by an algebraic relation. Then, after introducing the standard Zerilli master function, the gravitational perturbation equation becomes~\cite{Zerilli:1970se,Zerilli:1971wd}
\begin{align}
\frac{d^{2}Z}{dx^{2}}+\left[\omega^{2}-V_{\rm polar}(x)\right]Z=0~,
\end{align}
where the effective potential for polar perturbation reads~\cite{Toshmatov:2016bsb},
\begin{align}
V_{\rm polar}&=\frac{g(r)}{2r^{2}K}\Bigg[2r\Big\{1+2rf'K'+g\left(rK''-4K' \right)\Big\}+K\left(2q+12f-7rg'+r^{2}g''\right)\Bigg]~,
\end{align}
with $K=\frac{2r}{2(q+1)-2g+rg'}$ and $q=\frac{(\ell-1)(\ell+2)}{2}$.
These provide the details for the polar perturbation of braneworld BH. 

\end{widetext}

\bibliography{References}

\providecommand{\href}[2]{#2}\begingroup\raggedright\begin{thebibliography}{10}

\bibitem{Hawking:1975vcx}
S.~W. Hawking, ``{Particle Creation by Black Holes},''
  \href{http://dx.doi.org/10.1007/BF02345020}{{\em Commun. Math. Phys.}
  {\bfseries 43} (1975) 199--220}. [Erratum: Commun.Math.Phys. 46, 206 (1976)].

\bibitem{Mathur:2009hf}
S.~D. Mathur, ``{The Information paradox: A Pedagogical introduction},''
  \href{http://dx.doi.org/10.1088/0264-9381/26/22/224001}{{\em Class. Quant.
  Grav.} {\bfseries 26} (2009) 224001},
\href{http://arxiv.org/abs/0909.1038}{{\ttfamily arXiv:0909.1038 [hep-th]}}.

\bibitem{Polchinski:2016hrw}
J.~Polchinski, \href{http://dx.doi.org/10.1142/9789813149441_0006}{``{The Black
  Hole Information Problem},''} in {\em {Theoretical Advanced Study Institute
  in Elementary Particle Physics}: {New Frontiers in Fields and Strings}},
  pp.~353--397.
\newblock 2017.
\newblock \href{http://arxiv.org/abs/1609.04036}{{\ttfamily arXiv:1609.04036
  [hep-th]}}.

\bibitem{Chakraborty:2017pmn}
S.~Chakraborty and K.~Lochan, ``{Black Holes: Eliminating Information or
  Illuminating New Physics?},''
  \href{http://dx.doi.org/10.3390/universe3030055}{{\em Universe} {\bfseries 3}
  no.~3, (2017) 55}, \href{http://arxiv.org/abs/1702.07487}{{\ttfamily
  arXiv:1702.07487 [gr-qc]}}.

\bibitem{Raju:2020smc}
S.~Raju, ``{Lessons from the information paradox},''
  \href{http://dx.doi.org/10.1016/j.physrep.2021.10.001}{{\em Phys. Rept.}
  {\bfseries 943} (2022) 2187},
  \href{http://arxiv.org/abs/2012.05770}{{\ttfamily arXiv:2012.05770
  [hep-th]}}.

\bibitem{Mathur:2005zp}
S.~D. Mathur, ``{The Fuzzball proposal for black holes: An Elementary
  review},'' \href{http://dx.doi.org/10.1002/prop.200410203}{{\em Fortsch.
  Phys.} {\bfseries 53} (2005) 793--827},
\href{http://arxiv.org/abs/hep-th/0502050}{{\ttfamily arXiv:hep-th/0502050
  [hep-th]}}.

\bibitem{Mathur:2008nj}
S.~D. Mathur, ``{Fuzzballs and the information paradox: A Summary and
  conjectures},''
\href{http://arxiv.org/abs/0810.4525}{{\ttfamily arXiv:0810.4525 [hep-th]}}.

\bibitem{Visser:2009pw}
M.~Visser, C.~Barcelo, S.~Liberati, and S.~Sonego, ``{Small, dark, and heavy:
  But is it a black hole?},'' \href{http://arxiv.org/abs/0902.0346}{{\ttfamily
  arXiv:0902.0346 [gr-qc]}}.
[PoSBHGRS,010(2008)].

\bibitem{Barcelo:2010xk}
C.~Barcelo, S.~Liberati, S.~Sonego, and M.~Visser, ``{Hawking-like radiation
  from evolving black holes and compact horizonless objects},''
  \href{http://dx.doi.org/10.1007/JHEP02(2011)003}{{\em JHEP} {\bfseries 02}
  (2011) 003},
\href{http://arxiv.org/abs/1011.5911}{{\ttfamily arXiv:1011.5911 [gr-qc]}}.

\bibitem{Giddings:2017jts}
S.~B. Giddings, ``{Astronomical tests for quantum black hole structure},''
\href{http://arxiv.org/abs/1703.03387}{{\ttfamily arXiv:1703.03387 [gr-qc]}}.

\bibitem{Carballo-Rubio:2018jzw}
R.~Carballo-Rubio, F.~Di~Filippo, S.~Liberati, and M.~Visser,
  ``{Phenomenological aspects of black holes beyond general relativity},''
  \href{http://dx.doi.org/10.1103/PhysRevD.98.124009}{{\em Phys. Rev.}
  {\bfseries D98} no.~12, (2018) 124009},
\href{http://arxiv.org/abs/1809.08238}{{\ttfamily arXiv:1809.08238 [gr-qc]}}.

\bibitem{Koshelev:2017bxd}
A.~S. Koshelev and A.~Mazumdar, ``{Do massive compact objects without event
  horizon exist in infinite derivative gravity?},''
  \href{http://dx.doi.org/10.1103/PhysRevD.96.084069}{{\em Phys. Rev.}
  {\bfseries D96} no.~8, (2017) 084069},
\href{http://arxiv.org/abs/1707.00273}{{\ttfamily arXiv:1707.00273 [gr-qc]}}.

\bibitem{Buoninfante:2018rlq}
L.~Buoninfante, A.~S. Koshelev, G.~Lambiase, J.~Marto, and A.~Mazumdar,
  ``{Conformally-flat, non-singular static metric in infinite derivative
  gravity},'' \href{http://dx.doi.org/10.1088/1475-7516/2018/06/014}{{\em JCAP}
  {\bfseries 1806} no.~06, (2018) 014},
\href{http://arxiv.org/abs/1804.08195}{{\ttfamily arXiv:1804.08195 [gr-qc]}}.

\bibitem{Buoninfante:2018xiw}
L.~Buoninfante, A.~S. Koshelev, G.~Lambiase, and A.~Mazumdar, ``{Classical
  properties of non-local, ghost- and singularity-free gravity},''
  \href{http://dx.doi.org/10.1088/1475-7516/2018/09/034}{{\em JCAP} {\bfseries
  09} (2018) 034}, \href{http://arxiv.org/abs/1802.00399}{{\ttfamily
  arXiv:1802.00399 [gr-qc]}}.

\bibitem{Buoninfante:2019teo}
L.~Buoninfante, A.~Mazumdar, and J.~Peng, ``{Nonlocality amplifies echoes},''
  \href{http://dx.doi.org/10.1103/PhysRevD.100.104059}{{\em Phys. Rev. D}
  {\bfseries 100} no.~10, (2019) 104059},
  \href{http://arxiv.org/abs/1906.03624}{{\ttfamily arXiv:1906.03624 [gr-qc]}}.

\bibitem{Cardoso:2016oxy}
V.~Cardoso, S.~Hopper, C.~F.~B. Macedo, C.~Palenzuela, and P.~Pani,
  ``{Gravitational-wave signatures of exotic compact objects and of quantum
  corrections at the horizon scale},''
  \href{http://dx.doi.org/10.1103/PhysRevD.94.084031}{{\em Phys. Rev.}
  {\bfseries D94} no.~8, (2016) 084031},
\href{http://arxiv.org/abs/1608.08637}{{\ttfamily arXiv:1608.08637 [gr-qc]}}.

\bibitem{Cardoso:2016rao}
V.~Cardoso, E.~Franzin, and P.~Pani, ``{Is the gravitational-wave ringdown a
  probe of the event horizon?},''
  \href{http://dx.doi.org/10.1103/PhysRevLett.116.171101}{{\em Phys. Rev.
  Lett.} {\bfseries 116} no.~17, (2016) 171101},
\href{http://arxiv.org/abs/1602.07309}{{\ttfamily arXiv:1602.07309 [gr-qc]}}.

\bibitem{Abedi:2016hgu}
J.~Abedi, H.~Dykaar, and N.~Afshordi, ``{Echoes from the Abyss: Tentative
  evidence for Planck-scale structure at black hole horizons},''
  \href{http://dx.doi.org/10.1103/PhysRevD.96.082004}{{\em Phys. Rev.}
  {\bfseries D96} no.~8, (2017) 082004},
\href{http://arxiv.org/abs/1612.00266}{{\ttfamily arXiv:1612.00266 [gr-qc]}}.

\bibitem{Holdom:2016nek}
B.~Holdom and J.~Ren, ``{Not quite a black hole},''
  \href{http://dx.doi.org/10.1103/PhysRevD.95.084034}{{\em Phys. Rev. D}
  {\bfseries 95} no.~8, (2017) 084034},
  \href{http://arxiv.org/abs/1612.04889}{{\ttfamily arXiv:1612.04889 [gr-qc]}}.

\bibitem{Abedi:2020ujo}
J.~Abedi, N.~Afshordi, N.~Oshita, and Q.~Wang, ``{Quantum Black Holes in the
  Sky},'' \href{http://dx.doi.org/10.3390/universe6030043}{{\em Universe}
  {\bfseries 6} no.~3, (2020) 43},
  \href{http://arxiv.org/abs/2001.09553}{{\ttfamily arXiv:2001.09553 [gr-qc]}}.

\bibitem{Maggio:2020jml}
E.~Maggio, L.~Buoninfante, A.~Mazumdar, and P.~Pani, ``{How does a dark compact
  object ringdown?},''
  \href{http://dx.doi.org/10.1103/PhysRevD.102.064053}{{\em Phys. Rev. D}
  {\bfseries 102} no.~6, (2020) 064053},
  \href{http://arxiv.org/abs/2006.14628}{{\ttfamily arXiv:2006.14628 [gr-qc]}}.

\bibitem{Maggio:2021ans}
E.~Maggio, P.~Pani, and G.~Raposo, ``{Testing the nature of dark compact
  objects with gravitational waves},''
  \href{http://arxiv.org/abs/2105.06410}{{\ttfamily arXiv:2105.06410 [gr-qc]}}.

\bibitem{Addazi:2021xuf}
A.~Addazi {\em et~al.}, ``{Quantum gravity phenomenology at the dawn of the
  multi-messenger era -- A review},''
  \href{http://arxiv.org/abs/2111.05659}{{\ttfamily arXiv:2111.05659
  [hep-ph]}}.

\bibitem{Burgess:2018pmm}
C.~P. Burgess, R.~Plestid, and M.~Rummel, ``{Effective Field Theory of Black
  Hole Echoes},'' \href{http://dx.doi.org/10.1007/JHEP09(2018)113}{{\em JHEP}
  {\bfseries 09} (2018) 113},
\href{http://arxiv.org/abs/1808.00847}{{\ttfamily arXiv:1808.00847 [gr-qc]}}.

\bibitem{Oshita:2019sat}
N.~Oshita, Q.~Wang, and N.~Afshordi, ``{On Reflectivity of Quantum Black Hole
  Horizons},'' \href{http://dx.doi.org/10.1088/1475-7516/2020/04/016}{{\em
  JCAP} {\bfseries 04} (2020) 016},
  \href{http://arxiv.org/abs/1905.00464}{{\ttfamily arXiv:1905.00464
  [hep-th]}}.

\bibitem{Damour:1982}
T.~{Damour}, ``{Surface Effects in Black-Hole Physics},'' in {\em Marcel
  Grossmann Meeting: General Relativity}, p.~587.
\newblock Jan., 1982.

\bibitem{Thorne:1986iy}
K.~S. Thorne, R.~Price, and D.~Macdonald, eds., {\em {BLACK HOLES: THE MEMBRANE
  PARADIGM}}.
\newblock 1986.

\bibitem{Price:1986yy}
R.~Price and K.~Thorne, ``{Membrane Viewpoint on Black Holes: Properties and
  Evolution of the Stretched Horizon},''
  \href{http://dx.doi.org/10.1103/PhysRevD.33.915}{{\em Phys. Rev. D}
  {\bfseries 33} (1986) 915--941}.

\bibitem{Maartens:2003tw}
R.~Maartens, ``{Brane world gravity},''
  \href{http://dx.doi.org/10.12942/lrr-2004-7}{{\em Living Rev. Rel.}
  {\bfseries 7} (2004) 7}, \href{http://arxiv.org/abs/gr-qc/0312059}{{\ttfamily
  arXiv:gr-qc/0312059}}.

\bibitem{Csaki:2004ay}
C.~Csaki, ``{TASI lectures on extra dimensions and branes},'' in {\em
  {Theoretical Advanced Study Institute in Elementary Particle Physics (TASI
  2002): Particle Physics and Cosmology: The Quest for Physics Beyond the
  Standard Model(s)}}, pp.~605--698.
\newblock 4, 2004.
\newblock \href{http://arxiv.org/abs/hep-ph/0404096}{{\ttfamily
  arXiv:hep-ph/0404096}}.

\bibitem{Perez-Lorenzana:2005fzz}
A.~Perez-Lorenzana, ``{An Introduction to extra dimensions},''
  \href{http://dx.doi.org/10.1088/1742-6596/18/1/006}{{\em J. Phys. Conf. Ser.}
  {\bfseries 18} (2005) 224--269},
  \href{http://arxiv.org/abs/hep-ph/0503177}{{\ttfamily arXiv:hep-ph/0503177}}.

\bibitem{Randall:1999ee}
L.~Randall and R.~Sundrum, ``{A Large mass hierarchy from a small extra
  dimension},'' \href{http://dx.doi.org/10.1103/PhysRevLett.83.3370}{{\em Phys.
  Rev. Lett.} {\bfseries 83} (1999) 3370--3373},
  \href{http://arxiv.org/abs/hep-ph/9905221}{{\ttfamily arXiv:hep-ph/9905221}}.

\bibitem{Arkani-Hamed:1998jmv}
N.~Arkani-Hamed, S.~Dimopoulos, and G.~R. Dvali, ``{The Hierarchy problem and
  new dimensions at a millimeter},''
  \href{http://dx.doi.org/10.1016/S0370-2693(98)00466-3}{{\em Phys. Lett. B}
  {\bfseries 429} (1998) 263--272},
  \href{http://arxiv.org/abs/hep-ph/9803315}{{\ttfamily arXiv:hep-ph/9803315}}.

\bibitem{Harko:2004ui}
T.~Harko and M.~K. Mak, ``{Vacuum solutions of the gravitational field
  equations in the brane world model},''
  \href{http://dx.doi.org/10.1103/PhysRevD.69.064020}{{\em Phys. Rev. D}
  {\bfseries 69} (2004) 064020},
  \href{http://arxiv.org/abs/gr-qc/0401049}{{\ttfamily arXiv:gr-qc/0401049}}.

\bibitem{Dadhich:2000am}
N.~Dadhich, R.~Maartens, P.~Papadopoulos, and V.~Rezania, ``{Black holes on the
  brane},'' \href{http://dx.doi.org/10.1016/S0370-2693(00)00798-X}{{\em Phys.
  Lett. B} {\bfseries 487} (2000) 1--6},
  \href{http://arxiv.org/abs/hep-th/0003061}{{\ttfamily arXiv:hep-th/0003061}}.

\bibitem{Chamblin:1999by}
A.~Chamblin, S.~W. Hawking, and H.~S. Reall, ``{Brane world black holes},''
  \href{http://dx.doi.org/10.1103/PhysRevD.61.065007}{{\em Phys. Rev. D}
  {\bfseries 61} (2000) 065007},
  \href{http://arxiv.org/abs/hep-th/9909205}{{\ttfamily arXiv:hep-th/9909205}}.

\bibitem{Chamblin:2000ra}
A.~Chamblin, H.~S. Reall, H.-a. Shinkai, and T.~Shiromizu, ``{Charged brane
  world black holes},''
  \href{http://dx.doi.org/10.1103/PhysRevD.63.064015}{{\em Phys. Rev. D}
  {\bfseries 63} (2001) 064015},
  \href{http://arxiv.org/abs/hep-th/0008177}{{\ttfamily arXiv:hep-th/0008177}}.

\bibitem{Emparan:1999wa}
R.~Emparan, G.~T. Horowitz, and R.~C. Myers, ``{Exact description of black
  holes on branes},''
  \href{http://dx.doi.org/10.1088/1126-6708/2000/01/007}{{\em JHEP} {\bfseries
  01} (2000) 007}, \href{http://arxiv.org/abs/hep-th/9911043}{{\ttfamily
  arXiv:hep-th/9911043}}.

\bibitem{Conklin:2017lwb}
R.~S. Conklin, B.~Holdom, and J.~Ren, ``{Gravitational wave echoes through new
  windows},'' \href{http://dx.doi.org/10.1103/PhysRevD.98.044021}{{\em Phys.
  Rev. D} {\bfseries 98} no.~4, (2018) 044021},
  \href{http://arxiv.org/abs/1712.06517}{{\ttfamily arXiv:1712.06517 [gr-qc]}}.

\bibitem{Cardoso:2017cqb}
V.~Cardoso and P.~Pani, ``{Tests for the existence of black holes through
  gravitational wave echoes},''
  \href{http://dx.doi.org/10.1038/s41550-017-0225-y}{{\em Nat. Astron.}
  {\bfseries 1} no.~9, (2017) 586--591},
\href{http://arxiv.org/abs/1709.01525}{{\ttfamily arXiv:1709.01525 [gr-qc]}}.

\bibitem{Cardoso:2019rvt}
V.~Cardoso and P.~Pani, ``{Testing the nature of dark compact objects: a status
  report},'' \href{http://dx.doi.org/10.1007/s41114-019-0020-4}{{\em Living
  Rev. Rel.} {\bfseries 22} no.~1, (2019) 4},
  \href{http://arxiv.org/abs/1904.05363}{{\ttfamily arXiv:1904.05363 [gr-qc]}}.

\bibitem{Chen:2020htz}
B.~Chen, Q.~Wang, and Y.~Chen, ``{Tidal response and near-horizon boundary
  conditions for spinning exotic compact objects},''
  \href{http://dx.doi.org/10.1103/PhysRevD.103.104054}{{\em Phys. Rev. D}
  {\bfseries 103} no.~10, (2021) 104054},
  \href{http://arxiv.org/abs/2012.10842}{{\ttfamily arXiv:2012.10842 [gr-qc]}}.

\bibitem{Xin:2021zir}
S.~Xin, B.~Chen, R.~K.~L. Lo, L.~Sun, W.-B. Han, X.~Zhong, M.~Srivastava,
  S.~Ma, Q.~Wang, and Y.~Chen, ``{Gravitational-wave echoes from spinning
  exotic compact objects: Numerical waveforms from the Teukolsky equation},''
  \href{http://dx.doi.org/10.1103/PhysRevD.104.104005}{{\em Phys. Rev. D}
  {\bfseries 104} no.~10, (2021) 104005},
  \href{http://arxiv.org/abs/2105.12313}{{\ttfamily arXiv:2105.12313 [gr-qc]}}.

\bibitem{Darmois1927}
G.~Darmois, {\em Les équations de la gravitation einsteinienne}.
\newblock Gauthier-Villars, 1927.
\newblock \url{http://eudml.org/doc/192556}.

\bibitem{Israel:1966rt}
W.~Israel, ``{Singular hypersurfaces and thin shells in general relativity},''
  \href{http://dx.doi.org/10.1007/BF02710419, 10.1007/BF02712210}{{\em Nuovo
  Cim.} {\bfseries B44S10} (1966) 1}.
[Nuovo Cim.B44,1(1966)].

\bibitem{VisserBook}
M.~Visser, {\em {Lorentzian wormholes: From Einstein to Hawking}}.
\newblock AIP, Woodbury, USA,
1996.
\newblock

\bibitem{Regge:1957td}
T.~Regge and J.~A. Wheeler, ``{Stability of a Schwarzschild singularity},''
  \href{http://dx.doi.org/10.1103/PhysRev.108.1063}{{\em Phys. Rev.} {\bfseries
  108} (1957) 1063--1069}.

\bibitem{Pani:2009ss}
P.~Pani, E.~Berti, V.~Cardoso, Y.~Chen, and R.~Norte, ``{Gravitational wave
  signatures of the absence of an event horizon. I. Nonradial oscillations of a
  thin-shell gravastar},''
  \href{http://dx.doi.org/10.1103/PhysRevD.80.124047}{{\em Phys. Rev.}
  {\bfseries D80} (2009) 124047},
\href{http://arxiv.org/abs/0909.0287}{{\ttfamily arXiv:0909.0287 [gr-qc]}}.

\bibitem{TheLIGOScientific:2016src}
{\bfseries LIGO Scientific, Virgo} Collaboration, B.~Abbott {\em et~al.},
  ``{Tests of general relativity with GW150914},''
  \href{http://dx.doi.org/10.1103/PhysRevLett.116.221101}{{\em Phys. Rev.
  Lett.} {\bfseries 116} no.~22, (2016) 221101},
  \href{http://arxiv.org/abs/1602.03841}{{\ttfamily arXiv:1602.03841 [gr-qc]}}.
  [Erratum: Phys.Rev.Lett. 121, 129902 (2018)].

\bibitem{Ghosh:2021mrv}
A.~Ghosh, R.~Brito, and A.~Buonanno, ``{Constraints on quasinormal-mode
  frequencies with LIGO-Virgo binary\textendash{}black-hole observations},''
  \href{http://dx.doi.org/10.1103/PhysRevD.103.124041}{{\em Phys. Rev. D}
  {\bfseries 103} no.~12, (2021) 124041},
  \href{http://arxiv.org/abs/2104.01906}{{\ttfamily arXiv:2104.01906 [gr-qc]}}.

\bibitem{Cardoso:2017cfl}
V.~Cardoso, E.~Franzin, A.~Maselli, P.~Pani, and G.~Raposo, ``{Testing
  strong-field gravity with tidal Love numbers},''
  \href{http://dx.doi.org/10.1103/PhysRevD.95.084014}{{\em Phys. Rev.}
  {\bfseries D95} no.~8, (2017) 084014},
\href{http://arxiv.org/abs/1701.01116}{{\ttfamily arXiv:1701.01116 [gr-qc]}}.

\bibitem{Garfinkle:2022dnm}
D.~Garfinkle, ``{Gravitational wave memory and the wave equation},''
  \href{http://arxiv.org/abs/2201.05543}{{\ttfamily arXiv:2201.05543 [gr-qc]}}.

\bibitem{Dey:2020lhq}
R.~Dey, S.~Chakraborty, and N.~Afshordi, ``{Echoes from braneworld black
  holes},'' \href{http://dx.doi.org/10.1103/PhysRevD.101.104014}{{\em Phys.
  Rev. D} {\bfseries 101} no.~10, (2020) 104014},
  \href{http://arxiv.org/abs/2001.01301}{{\ttfamily arXiv:2001.01301 [gr-qc]}}.

\bibitem{Mishra:2021waw}
A.~K. Mishra, A.~Ghosh, and S.~Chakraborty, ``{Constraining extra dimensions
  using observations of black hole quasi-normal modes},''
  \href{http://arxiv.org/abs/2106.05558}{{\ttfamily arXiv:2106.05558 [gr-qc]}}.

\bibitem{Banerjee:2021aln}
I.~Banerjee, S.~Chakraborty, and S.~SenGupta, ``{Looking for extra dimensions
  in the observed quasi-periodic oscillations of black holes},''
  \href{http://arxiv.org/abs/2105.06636}{{\ttfamily arXiv:2105.06636 [gr-qc]}}.

\bibitem{Chakraborty:2021gdf}
S.~Chakraborty, S.~Datta, and S.~Sau, ``{Tidal heating of black holes and
  exotic compact objects on the brane},''
  \href{http://dx.doi.org/10.1103/PhysRevD.104.104001}{{\em Phys. Rev. D}
  {\bfseries 104} no.~10, (2021) 104001},
  \href{http://arxiv.org/abs/2103.12430}{{\ttfamily arXiv:2103.12430 [gr-qc]}}.

\bibitem{Dey:2020pth}
R.~Dey, S.~Biswas, and S.~Chakraborty, ``{Ergoregion instability and echoes for
  braneworld black holes: Scalar, electromagnetic, and gravitational
  perturbations},'' \href{http://dx.doi.org/10.1103/PhysRevD.103.084019}{{\em
  Phys. Rev. D} {\bfseries 103} no.~8, (2021) 084019},
  \href{http://arxiv.org/abs/2010.07966}{{\ttfamily arXiv:2010.07966 [gr-qc]}}.

\bibitem{Banerjee:2019nnj}
I.~Banerjee, S.~Chakraborty, and S.~SenGupta, ``{Silhouette of M87*: A New
  Window to Peek into the World of Hidden Dimensions},''
  \href{http://dx.doi.org/10.1103/PhysRevD.101.041301}{{\em Phys. Rev. D}
  {\bfseries 101} no.~4, (2020) 041301},
  \href{http://arxiv.org/abs/1909.09385}{{\ttfamily arXiv:1909.09385 [gr-qc]}}.

\bibitem{Chakraborty:2017qve}
S.~Chakraborty, K.~Chakravarti, S.~Bose, and S.~SenGupta, ``{Signatures of
  extra dimensions in gravitational waves from black hole quasinormal modes},''
  \href{http://dx.doi.org/10.1103/PhysRevD.97.104053}{{\em Phys. Rev. D}
  {\bfseries 97} no.~10, (2018) 104053},
  \href{http://arxiv.org/abs/1710.05188}{{\ttfamily arXiv:1710.05188 [gr-qc]}}.

\bibitem{Chakravarti:2019aup}
K.~Chakravarti, S.~Chakraborty, K.~S. Phukon, S.~Bose, and S.~SenGupta,
  ``{Constraining extra-spatial dimensions with observations of GW170817},''
  \href{http://dx.doi.org/10.1088/1361-6382/ab8355}{{\em Class. Quant. Grav.}
  {\bfseries 37} no.~10, (2020) 105004},
  \href{http://arxiv.org/abs/1903.10159}{{\ttfamily arXiv:1903.10159 [gr-qc]}}.

\bibitem{Fabbri:2007kr}
A.~Fabbri and G.~P. Procopio, ``{Quantum effects in black holes from the
  Schwarzschild black string?},''
  \href{http://dx.doi.org/10.1088/0264-9381/24/22/003}{{\em Class. Quant.
  Grav.} {\bfseries 24} (2007) 5371--5382},
  \href{http://arxiv.org/abs/0704.3728}{{\ttfamily arXiv:0704.3728 [hep-th]}}.

\bibitem{Kanno:2003au}
S.~Kanno and J.~Soda, ``{Rotating black string and effective Teukolsky equation
  in brane world},'' \href{http://dx.doi.org/10.1088/0264-9381/21/7/012}{{\em
  Class. Quant. Grav.} {\bfseries 21} (2004) 1915--1924},
  \href{http://arxiv.org/abs/gr-qc/0311074}{{\ttfamily arXiv:gr-qc/0311074}}.

\bibitem{Kanno:2003sc}
S.~Kanno and J.~Soda, ``{Effective Teukolsky equation on the brane},'' in {\em
  {6th RESCEU International Symposium on Frontier in Astroparticle Physics and
  Cosmology}}.
\newblock 12, 2003.
\newblock \href{http://arxiv.org/abs/gr-qc/0312098}{{\ttfamily
  arXiv:gr-qc/0312098}}.

\bibitem{Toshmatov:2016bsb}
B.~Toshmatov, Z.~Stuchl\'\i{}k, J.~Schee, and B.~Ahmedov, ``{Quasinormal
  frequencies of black hole in the braneworld},''
  \href{http://dx.doi.org/10.1103/PhysRevD.93.124017}{{\em Phys. Rev. D}
  {\bfseries 93} no.~12, (2016) 124017},
  \href{http://arxiv.org/abs/1605.02058}{{\ttfamily arXiv:1605.02058 [gr-qc]}}.

\bibitem{Rahman:2021kwb}
M.~Rahman and A.~Bhattacharyya, ``{Ringdown of charged compact objects using
  membrane paradigm},''
  \href{http://dx.doi.org/10.1103/PhysRevD.104.044045}{{\em Phys. Rev. D}
  {\bfseries 104} no.~4, (2021) 044045},
  \href{http://arxiv.org/abs/2104.00074}{{\ttfamily arXiv:2104.00074 [gr-qc]}}.

\bibitem{Wang:2019rcf}
Q.~Wang, N.~Oshita, and N.~Afshordi, ``{Echoes from Quantum Black Holes},''
  \href{http://dx.doi.org/10.1103/PhysRevD.101.024031}{{\em Phys. Rev. D}
  {\bfseries 101} no.~2, (2020) 024031},
  \href{http://arxiv.org/abs/1905.00446}{{\ttfamily arXiv:1905.00446 [gr-qc]}}.

\bibitem{Abedi:2021tti}
J.~Abedi, L.~F.~L. Micchi, and N.~Afshordi, ``{GW190521: First Measurement of
  Stimulated Hawking Radiation from Black Holes},''
  \href{http://arxiv.org/abs/2201.00047}{{\ttfamily arXiv:2201.00047 [gr-qc]}}.

\bibitem{Zerilli:1970se}
F.~J. Zerilli, ``{Effective potential for even parity Regge-Wheeler
  gravitational perturbation equations},''
  \href{http://dx.doi.org/10.1103/PhysRevLett.24.737}{{\em Phys. Rev. Lett.}
  {\bfseries 24} (1970) 737--738}.

\bibitem{Zerilli:1971wd}
F.~Zerilli, ``{Gravitational field of a particle falling in a schwarzschild
  geometry analyzed in tensor harmonics},''
\href{http://dx.doi.org/10.1103/PhysRevD.2.2141}{{\em Phys. Rev.} {\bfseries
  D2} (1970) 2141--2160}.

\end{thebibliography}\endgroup

\bibliographystyle{./utphys1}
\end{document}